\documentclass[twocolumn]{aa}
\usepackage{graphicx}
\usepackage{epsfig,graphicx,rotating,portland,fancyheadings,longtable}
\usepackage{multirow}
\usepackage{txfonts}

\begin{document}
\title{VLT/X-shooter spectroscopy of the GRB\,120327A afterglow%
  \thanks{Based on observations collected at the European Southern
    Observatory, ESO, the VLT/Kueyen telescope, Paranal, Chile,
    proposal code: 088.A-0051.}}

\author{V. D'Elia$^{1,2}$, J. P. U. Fynbo$^{3}$, P. Goldoni$^{4}$,
  S. Covino$^{5}$, A. de Ugarte Postigo$^{6,3}$, C. Ledoux$^{7}$,
  F. Calura$^{8}$, J. Gorosabel$^{6,9,10}$, D. Malesani$^{3}$,
  F. Matteucci$^{11}$ R. S\'anchez-Ram\'irez$^{6}$,
  S. Savaglio$^{12}$, A. J. Castro-Tirado$^{6}$, O. E. Hartoog$^{13}$,
  L. Kaper$^{13}$ T. Mu\~noz-Darias$^{14,15}$, E. Pian$^{16}$, S. Piranomonte$^{1}$
  G. Tagliaferri$^{5}$, N. Tanvir$^{17}$ S. D. Vergani$^{18}$,
  D. J. Watson$^{3}$, D. Xu$^{3}$
}

\institute
{$^1$INAF-Osservatorio Astronomico di Roma, via Frascati 33, I-00040 Monteporzio Catone, Italy\\
$^2$ASI-Science Data Center, Via del Politecnico snc, I-00133 Rome, Italy\\
$^3$Dark Cosmology Centre, Niels Bohr Institute, University of Copenhagen, Juliane Maries Vej 30, 2100 Copenhagen, Denmark  \\
$^4$APC, Astroparticule et Cosmologie, Univ. Paris Diderot, CNRS/IN2P3, CEA/Irfu, Obs. de Paris, Sorbonne Paris Cité, 10, Rue Alice Domon et Léonie Duquet, 75205, Paris Cedex 13, France
$^{5}$INAF-Osservatorio Astronomico di Brera, via E. Bianchi 46, 23807 Merate (LC), Italy\\
$^{6}$Instituto de Astrof\'isica de Andaluc\'ia (IAA-CSIC), Glorieta de la Astronomía s/n, 18008, Granada, Spain\\ 
$^{7}$European Southern Observatory, Alonso de Cordova 3107, Vitacura, Casilla 19001, Santiago 19, Chile \\
$^{8}$INAF, Osservatorio Astronomico di Bologna, via Ranzani 1 I-40127 Bologna, Italy \\
$^{9}$Unidad Asociada Grupo Ciencia Planetarias UPV/EHU-IAA/CSIC, Departamento de F\'isica Aplicada I, E.T.S. Ingenieria, 
Universidad del Pais Vasco UPV/EHU, Alameda de Urquijo s/n, E-48013, Bilbao, Spain\\
$^{10}$Ikerbasque, Basque Foundation for Science, Alameda de Urquijo 36-5, E-48008 Bilbao, Spain\\
$^{11}$INAF, Osservatorio Astronomico di Trieste, Via Tiepolo 11, 34143 Trieste, Italy \\
$^{12}$Max Planck Institute for Extraterrestrial Physics, 85748 Garching bei M\"unchen, Germany\\
$^{13}$Astronomical Institute ``Anton Pannekoek'', University of Amsterdam, Science Park 904, 1098 XH Amsterdam, The Netherlands \\
$^{14}$School of Physics and Astronomy University of Southampton, Southampton, Hampshire, SO17 1BJ, United Kingdom\\
$^{15}$University of Oxford, Department of Physics, Astrophysics, Keble Road, Oxford, OX1 3RH, United Kingdom
$^{16}$Scuola Normale Superiore, Piazza dei Cavalieri 7, 56126, Pisa, Italy\\
$^{17}$Department of Physics and Astronomy, University of Leicester, Leicester LE1 7RH, UK \\
$^{18}$GEPI, Observatoire de Paris, CNRS, Univ. Paris Diderot, 5 place Jules Janssen, 92190, Meudon, France\\
}

  \abstract 
  {} 
  {We present a study of the environment of the Swift long gamma-ray
    burst GRB\,120327A at $z\approx 2.8$ through optical spectroscopy
    of its afterglow.}
  {We analyzed medium-resolution, multi-epoch spectroscopic
    observations ($R \sim 7000\mbox{--}12000$, corresponding to $\sim
    15\mbox{--} 23$ km~s$^{-1}$, $\mbox{S/N} = 15 \mbox{--} 30$ and
    wavelength range 3000--25000~\AA) of the optical afterglow of
    GRB\,120327A, taken with X-shooter at the VLT $2.13$ and $27.65$ hr
    after the GRB trigger.}
  {The first epoch spectrum shows that the ISM in the GRB host galaxy
    at $z = 2.8145$ is extremely rich in absorption features, with
    three components contributing to the line profiles. The hydrogen
    column density associated with GRB\,120327A has $\log N_{\rm
      H}/{\rm cm}^{-2} = 22.01 \pm 0.09$, and the metallicity of the
    host galaxy is in the range $\mbox{[X/H]} = -1.3$ to $-1.1$. In
    addition to the ground state lines, we detect absorption features
    associated with excited states of \ion{C}{ii}, \ion{O}{i},
    \ion{Si}{ii}, \ion{Fe}{ii}, and \ion{Ni}{ii}, which we used to
    derive information on the distance between the host absorbing gas
    and the site of the GRB explosion. The variability of the 
    {\ion{Fe}{ii}~$\lambda$2396} excited line between the two epochs
    proves that these features are excited by the GRB UV
    flux. Moreover, the distance of component I is found to be $d_{\rm
      I}=200^{+100}_{-60}$~pc, while component II is located closer to
    the GRB, at $d_{\rm II}=100^{+40}_{-30}$~pc. These values are
    among the lowest found in GRBs. Component III does not show
    excited transitions, so it should be located farther away from the
    GRB.

    The presence of H$_2$ molecules is firmly established, with a
    molecular fraction $f$ in the range $f=4\times 10^{-7} \mbox{--}
    10^{-4}$. This particularly low value can be attributed to the
    small dust content.  This represents the third positive detection
    of molecules in a GRB environment.

}
{}

\authorrunning {D'Elia et al.}
\titlerunning {X-shooter spectroscopy of GRB\,120327A afterglow}

\maketitle
%

\section{Introduction}

For a few hours after their onset, long gamma-ray burst (GRB)
afterglows (see, e.g., M\'esz\'aros 2006 for a review) are the
brightest sources in the Universe. Their intense and featureless
continua in the X-ray to the radio band offer a superb opportunity to
study the gas and dust content of high redshift galaxies (see e.g.,
Vreeswijk et al. 2001; Vreeswijk et al. 2004; Berger et al. 2006;
Fynbo et al. 2006; Prochaska et al. 2007; Sparre et
al. 2013). Therefore, GRBs provide a complementary diagnostic tool to
study high-redshift environments, with respect to
Lyman-break galaxies (LBGs, see e.g. Steidel et al.  1999), $K$-band
selected galaxies (Savaglio et al. 2004; Martin et al. 2012; Kornei et
al. 2013), and galaxies that happen to be along the lines of sight to
bright background QSOs. In fact, LBG studies are biased towards high
UV-luminosity galaxies, and their inferred chemical abundances may not
be related to that of typical high-redshift galaxies.

Ultra-deep spectroscopic observations may allow us to study normal
galaxies, but only up to $z \sim 2$ (using the present generation of
8-m class telescopes). For example, Savaglio et al. (2004, 2005)
studied the inter-stella medium (ISM) of a sample of faint $K$-band
selected galaxies, finding abundances much higher than in QSOs
systems. This is because the line of sight to QSO probes mainly galaxy
halos, rather than their bulges or discs (Fynbo et al. 2008),
returning lower abundance values. GRBs represent a way to study the
central regions of high-redshift galaxies, without luminosity
biases. Abundances measured in GRB host galaxies vary from less than
$10^{-2}$ to solar values and are on average higher than those found
along QSO sightlines (see e.g., Fynbo et al. 2006; Prochaska et
al. 2007; Savaglio et al. 2012; Sparre et al. 2013). GRBs are expected
to originate in molecular clouds, where star formation
occurs. Nevertheless, absorption from ground-state and vibrationally
excited levels of the H$_2$ and other molecules is hardly observed in
GRB afterglow spectra (Vreeswijk et al. 2004, Tumlinson et al. 2007),
with the exception of GRB\,080607 (Prochaska et al. 2009, Sheffer et
al. 2009), GRB\,120815A (Kr\"uhler et al. 2013), and possibly
GRB\,060206 (Fynbo et al. 2006). The non-detection of molecules is
possibly a consequence of photo-dissociation by the intense UV flux
from the GRB afterglow or by the young, hot stellar populations
harboring the GRB host.

Furthermore, GRBs are the most variable and violent phenomena in the
Universe, while QSOs are persistent, albeit variable, sources. Thus,
the physical, dynamical and chemical status of the circumburst medium
in the star-forming region hosting GRB progenitors is altered in real
time by the explosive event, through shock waves and ionizing
photons. The transient nature of GRBs is manifested through
variability in the populations of fine structure and other excited
levels of the {\ion{O}{i}} atom and of ions such as {\ion{Fe}{ii}},
{\ion{Ni}{ii}}, {\ion{Si}{ii}} and {\ion{C}{ii}}. The population of
these excited levels, and therefore the observed column densities of
the corresponding absorption lines, change on timescales of minutes to
days (Vreeswijk et al. 2007, D'Elia et al. 2009a, De Cia et al. 2012,
Hartoog et al. 2013). This variation is not consistent with a pure
infrared excitation or collisional processes, but is due to UV-pumping
(Prochaska, Chen \& Bloom 2006, Vreeswijk et al. 2007). Monitoring
these variations and comparing them with predictions from
photo-excitation codes enables one to derive information on the
distance between the GRB and the absorbing material. This distance can
be estimated even when multi-epoch spectroscopy is not available,
under the assumption that UV pumping is the responsible mechanism for
the production of the excited lines (D'Elia et al. 2009b; 2010; 2011;
Ledoux et al. 2009).



The GRB environments are in general very complex. In fact, they are
constituted by several layers of gas which may be close to or far from
one another, have different relative velocities and physical
properties. Optical spectroscopy reflects this complexity, since the
absorption lines in several cases cannot be fitted with a single line
profile. The ISM complexity is particularly evident for GRB afterglows
observed at high resolution (Fiore et al. 2005; D'Elia et al. 2007;
Th\"one et al. 2008; Piranomonte et al. 2008).

Here we present the case of GRB\,120327A, observed by X-shooter on
2012 March 27 and 28. Our dataset is complemented by a GTC spectrum and
photometric data from several facilities. We derive the abundances by
comparing the column densities of hydrogen and metals, and evaluate
the GRB-absorber distance using the variation of the excited levels
and their ratios to the ground state ones, and the predictions of
photo-excitation codes. We also characterize the host galaxy using the
dust depletion pattern technique and chemical evolution models, derive
the extinction curve shape and search for H$_2$ molecules in the GRB
environment.

The paper is organized as follows. Section 2 summarizes the properties
of GRB\,120327A and results of observations previously reported in the
literature; Sect. 3 presents our observations and data reduction;
Sect. 4 is devoted to the analysis of the absorption features from the
host galaxy, using X-shooter data; Sect. 5 illustrates our main
results, which are discussed in Sect. 6. Finally, in Sect. 7 the
conclusions are drawn. We assume a cosmology with $H_0 = 70$ km
s$^{-1}$ Mpc$^{-1}$, $\Omega_{\rm m} = 0.3$, $\Omega_\Lambda =
0.7$. Hereafter, the notation [X/H] indicates the logarithmic
abundance of element X relative to solar values.

\section{GRB\,120327A}

GRB\,120327A was discovered by the Burst Alert Telescope (BAT)
instrument on board {\it Swift} (Gehrels et al. 2004) on 2012 March 27
at 02:55:16 UT (Sbarufatti et al. 2012). The BAT light curve shows 3
main spikes, peaking at $T{_0} + 2$ s, $T{_0} + 18$ s and $T{_0} + 37$
s, respectively, where $T_0$ is the burst detection time. The last one
is the main peak, and the duration for this burst in the $15-350$ keV
range is $T_{90} = 62.9 \pm 7.5$ s. The BAT spectrum is best fit by a
simple power-law model, with a photon index of $1.52\pm0.06$ (Krimm et
al. 2012).

The X-Ray Telescope (XRT) and Ultraviolet and the Optical Telescope
(UVOT) were repointed toward the GRB position $76$ s and $140$ s after
the trigger, respectively. Both instruments detected the
afterglow. The XRT light curve shows a typical steep-flat-normal
behavior with temporal indices of $\alpha_1=2.95$, $\alpha_2=0.62$,
and $\alpha_3=1.42$, with temporal breaks at $217$ s and $2540$ s
after the burst (Stroh et al. 2012).  The X-ray spectrum can be
modelled with an absorbed power law with a spectral index of $\sim 1$
and a column density of $N_{\rm H} \approx 1.5 \times 10^{22}$
cm$^{-2}$, taking into account the absorption by our Galaxy ($N_{\rm
  H,G} = 1.4 \times 10^{21}$ cm$^{-2}$; Kalberla et al. 2005).

Ground-based facilities were pointed to the GRB starting a few minutes
later, allowing the detection of the afterglow in the optical and near
infrared (Klotz et al. 2012; Smith \& Virgili 2012; LaCluyze et
al. 2012; Covino et al. 2012), and the measurement of a photometric
redshift of $z\sim3$ (Sudilowsky et al. 2012). Optical spectroscopy
followed later on, yielding a redshift of $2.813$ (Perley \& Tanvir
2012; Kr\"uhler et al. 2012; S\'anchez-Ram\'irez et
al. 2012). Finally, the afterglow was also detected in the radio band
at 34~GHz, $\approx 4.6$ days after the trigger (Hancock et al 2012).





\begin{table}[ht]
\caption{\bf X-shooter observations}
\centering
{\footnotesize
\smallskip
\begin{tabular}{|l|c|c|c|}
  \hline 
  Observation time (UT)  & Hours since GRB      & Exp. (s) &  S/N       \\
  \hline                          
  March 27, 05:03:56 (A) & 2.13                 & 600          &  $5-15$    \\
  \hline                          
  March 27, 05:15:51 (B) & 2.33                 & 600          &  $5-15$    \\
  \hline                            
  March 27, 05:27:35 (C) & 2.53                 & 600          &  $5-15$    \\
  \hline                          
  March 27, 05:39:16 (D) & 2.72                 & 600          &  $5-15$    \\
  \hline                         
  A+B+C+D = 1st epoch                & 2.47                 & 2400         &  $10-30$   \\
  \hline
  \hline
  March 28, 08:42:36     & 27.65                & $\sim$4200         &  $2-5$   \\
  \hline

\end{tabular}
}
\end{table}




\section{Observations and data reduction}

\subsection{X-shooter observations}

We observed the afterglow of GRB\,120327A twice with X-shooter
(D'Odorico et al. 2006; de Ugarte Postigo et al. 2010; Vernet et
al. 2011), a single-object medium resolution ($R=\lambda/\Delta\lambda =
4000\mbox{--}14000$) echelle spectrograph mounted at the VLT-UT2
telescope. The observations were carried out under the GTO programme
088.A-0051.

For both observations the slit width was set to 0.9\arcsec{} in the
VIS and NIR arms and 1.0\arcsec{} in the UVB arm. The UVB and VIS CCD
detectors were rebinned to $1 \times 2$ pixels (binned in the spectral
direction but not in the spatial one) to reduce the readout noise.
With this configuration, the nominal resolution is different for the three
arms: $R \sim 5100$, $8800$, $5300$ for the UVB, VIS, and NIR arms,
respectively.  The first observation began on 2012 March 27 at
05:03:56 UT, $\approx 2.13$~hr after the GRB trigger (the magnitude of
the afterglow at the time of this observation was $R = 18.84 \pm
0.09$, see Klotz et al. 2012). It consisted of 4 different exposures
of 600~s each for a total net exposure of 40~min (see Table 1). The
exposures were taken nodding along the slit with an offset of
$5\arcsec$ between exposures, following a standard ABBA pattern. We
started the second sequence the following night, on 2012 March 28, at
08:42:36 UT, using the same instrument configuration. Because of the
lower flux, the exposure time of every frame was 1200~s, though the
fourth exposure was stopped after $\sim 600$ s because of the incoming
twilight.

We processed the spectra using version 1.4.5 of the X-shooter data
reduction pipeline (Goldoni et al. 2006; Modigliani et al. 2010).  The
pipeline carries out the following steps: the raw frames were first
bias-subtracted, and cosmic ray hits were detected and removed using
the method developed by van Dokkum (2001). The frames were divided by
a master flat field obtained using day-time flat field exposures with
halogen lamps.  The orders were extracted and rectified in wavelength
space using a wavelength solution obtained from day-time calibration
frames.  The resulting rectified orders were shifted and co-added to
obtain the final two-dimensional spectrum. In the overlapping regions,
orders were merged by weighing them using the errors propagated during
the reduction process. From the resulting two-dimensional merged
spectrum, a one-dimensional spectrum with the corresponding error file
and bad pixel map was extracted at the source position.

To perform flux calibration we extracted a spectrum from a staring
observation of the flux standard star GD\,153 (Bohlin \& Gilliland
2004).  In this case we subtracted the sky emission lines using the
method developed by Kelson (2003).  This spectrum was divided by the
flux table of the same star from the CALSPEC HST
database\footnote{http://www.stsci.edu/hst/observatory/cdbs/calspec.html}
to produce the response function.  The response was then interpolated
where needed inside the atmospheric absorption bands in VIS and NIR
and applied to the spectrum of the source. No telluric correction was
applied, except while checking for variability of some {\ion{Fe}{ii}}
lines (see Sect. 4.7).

At a first glance the flux-calibrated VIS spectrum of the first
observation appeared to be anomalous with respect to the UVB and NIR
portions. Indeed in the overlapping region between the VIS and NIR arm,
the flux level of the VIS spectrum was about 10 times lower than the
flux level of the NIR spectrum. Moreover the VIS spectrum could be
fitted with a power law $F_\lambda \propto \lambda^{-3.3}$, inconsistent with
the UVB and NIR slopes. No such anomaly is present in the much weaker
spectra of the second observation.

As our flux calibration was carried out in a standard way, we
suspected a technical anomaly to be responsible for this effect. After
consulting with the X-shooter instrument scientist at Paranal,
Christophe Martayan, we concluded that this anomalous spectrum is the
result of a malfunction of the atmospheric dispersion corrector (ADC)
of the VIS arm. Indeed, starting from 2012 August, the X-shooter ADCs have
been disabled because of recurring problems of this kind. We checked that this
phenomenon does not impact the quality of the wavelength calibration
but only the flux level.  In the following we do not use this flux
calibrated VIS spectrum, but only the normalized one.

\subsection{GTC spectroscopic observations}

We also obtained long-slit spectroscopy of the GRB\,120327A afterglow
with the OSIRIS spectrograph (Cepa et al. 2003) mounted on the 10.4-m
Gran Telescopio de Canarias (GTC), located at the Roque de los Muchachos Observatory
(S\'anchez-Ram\'irez et al. 2012). We took $3 \times 400$~s spectra with the
R1000B grism, covering the range 3626--7883~\AA, 2.33~hr
after the BAT trigger. The resolution of our spectra is $R \sim 750$.

The data were reduced using standard procedures in IRAF and the flux
calibration is based on the spectrophotometric standard G191-B2B (Oke
1990). We have also scaled the combined spectrum to the magnitude of
the acquisition image ($r'$=18.5 $\pm$ 0.1 in the AB system, not
corrected for Galactic extinction of $E_{B-V} = 0.34$~mag) to compensate for
slit losses. 

\subsection{Photometric observations}

Late $r^\prime$-band images were taken with a total exposure time of
$8\times600$~s with the 2.0-m LT telescope, between 2013 April 23.15
and 23.21 UT. The observations failed to detect any source above a
$3\sigma$ limit of $r > 24.0$ (AB system, corrected for Galactic
extinction, and calibrated using the GTC data, see below) coincident
with the optical afterglow position (Kuin et al.  2012, Smith \&
Virgili 2012). Further $r^\prime$-band imaging was carried out with
the GTC equipped with OSIRIS. Twelve images with an exposure time of
50~s per frame were taken on 2012 May 26.096--26.11 UT.  The field was
calibrated with standard star PG\,1047 (Smith et al.  2002) observed
at an airmass similar to the one of the GRB field.  The combined image
reveals an object coincident with the afterglow position (Fig. 1),
which we propose to be the host galaxy, since the extrapolation of the
afterglow light curve at two months after the GRB explosion results in
negligible flux, and no contribution is expected in the rest-frame UV
from any underlying supernova.  Aperture photometry yields an AB
magnitude of $r^\prime=24.50\pm0.23$ for the host, including
calibration errors and not corrected for Galactic reddening.

Additional late $H$-band imaging was performed with the 3.5m Calar
Alto telescope equipped with $\Omega_{2000}$.  The observations were
done on 2012 May 1.0622--1.0902 UT, with a total exposure time of
1800~s.  The combined image reveals an object coincident with the GTC
detection, with $H = 20.09\pm0.32$ (Vega) not corrected for Galactic
reddening and including the calibration errors.  The $H$-band
photometric calibration was carried out with 468 field stars of the
2MASS catalogue.

In addition, the afterglow $I$-band decay was monitored by means of 14
images of 300~s each acquired with the IAC80 telescope from 1.19 to
2.63 hr after the GRB (Gorosabel et al. 2012, see also Appendix
  A).  The afterglow shows a smooth fading from $I = 17.32\pm0.05$ to
$I = 18.17\pm0.13$ (Vega).  The $I$-band calibration was carried out
using 21 stars from the USNO-B1.0 catalogue yielding a dispersion of
0.16 mag (not included in the error of the above two measurements). We
note that the last three IAC80 images are contemporaneous to the GTC
spectra (Fig. 1).

\begin{figure}
\centering
\includegraphics[angle=0,width=9cm]{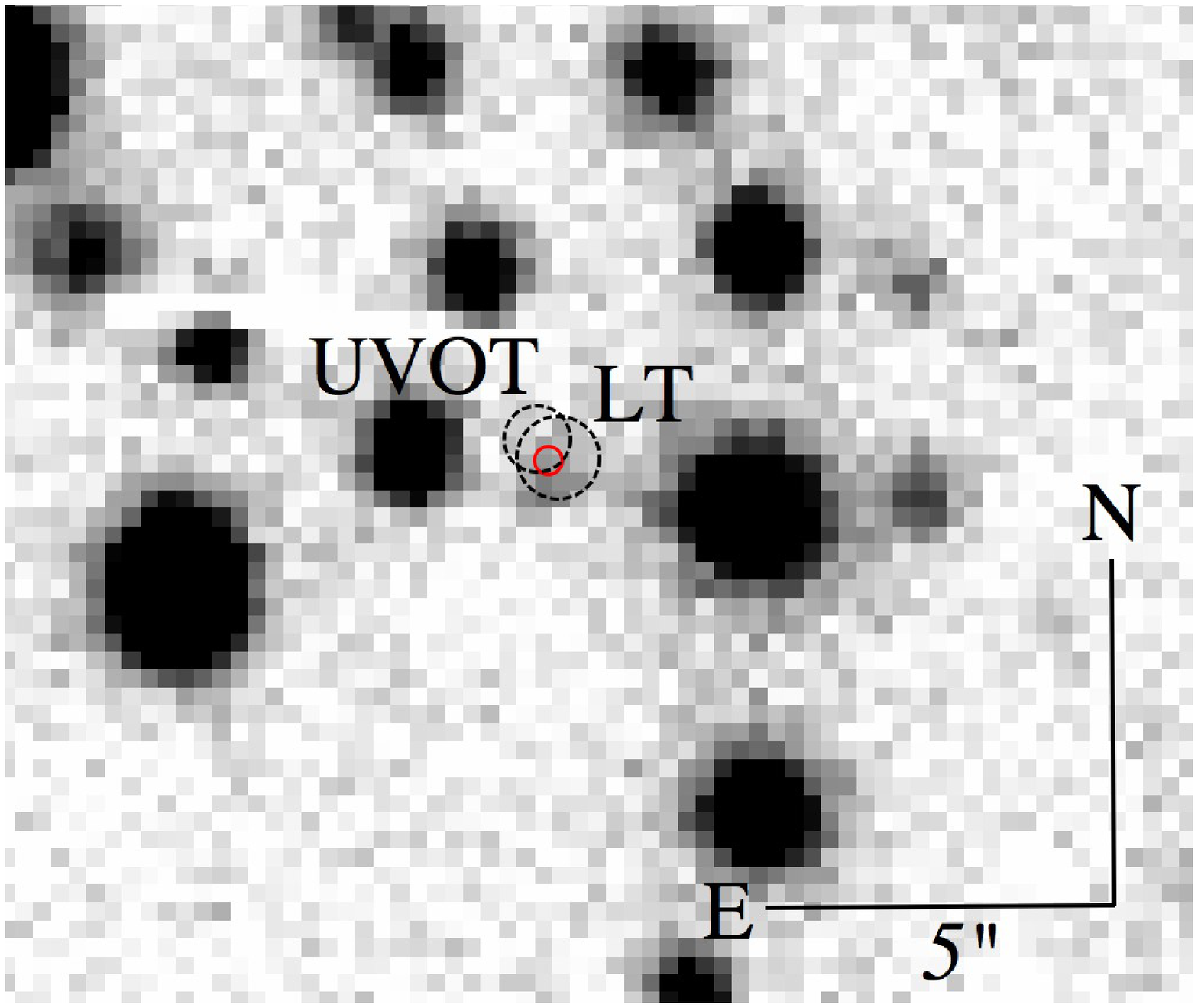}

\includegraphics[angle=-0,width=9.0cm]{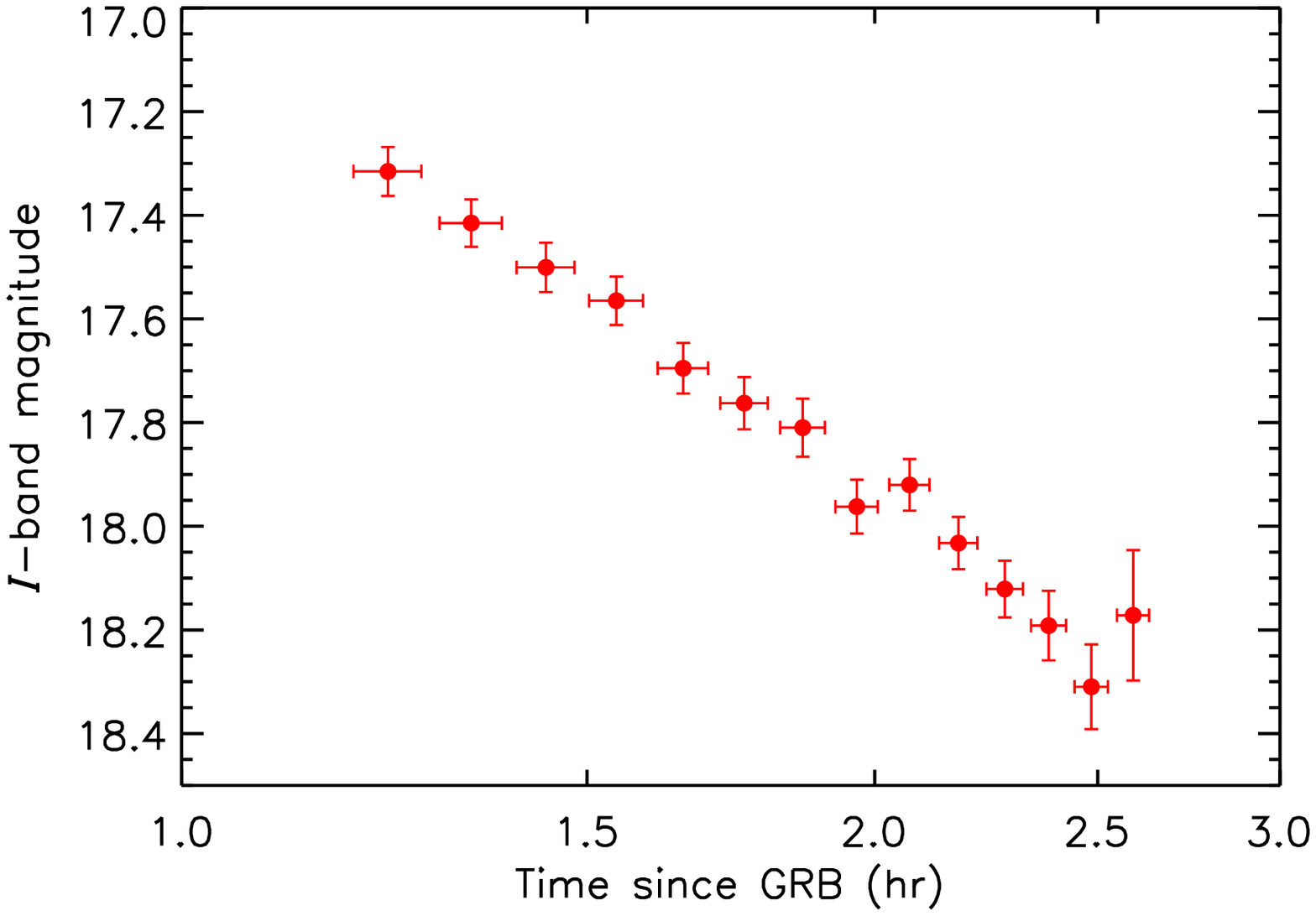}
\caption{Upper panel: combined $r^\prime$-band image result of the GTC
  late observations. The two long-dashed circles indicate the
  astrometric error circles of the UVOT (Kuin \& Sbarufatti 2012) and
  Liverpool Telescope (Smith \& Virgili 2012) positions reported for
  the optical afterglow. The small circle is the result of registering
  the acquisition image of the GTC afterglow spectrum on the deep late
  image taken with the same telescope. The three circles are
  consistent with a faint underlying object which we propose to be the
  host galaxy. Lower panel: The GRB\,120327A $I$-band light curve from
  IAC80.}
\end{figure}



\begin{table*}[ht]
\centering
  \caption{\bf Absorption line column densities for the three components of the main system.}
  {\footnotesize
    \smallskip
    \begin{tabular}{|lc|ccc|}
      \hline
      Atomic/ionic species                       & Observed transitions                      & I ($-41$ km s$^{-1}$) & II ($0$ km s$^{-1}$)& III ($+35$ km s$^{-1}$)  \\
      \hline
      \ion{H}{i}  $^2S_{1/2}$         & Ly$\alpha$, Ly$\beta$, Ly$\gamma$, Ly$\delta$& -                    & $22.01\pm0.09$     &  -\\
      \hline
      \ion{C}{ii} $^2P^{0}_{1/2}$      &  $\lambda$1036, $\lambda$1334             & & (S)$^{1}$       &       \\
      \hline
      \ion{C}{ii} $^2P^{0}_{3/2}$ (1s) &  $\lambda$1037, $\lambda$1335             & $14.67 \pm 0.04$    &$14.05 \pm 0.10$    &  $< 13.0$                     \\
      \hline
      \ion{C}{iv} $^2S_{1/2}$          &  $\lambda$1548, $\lambda$1550              &                    & (S)   &                        \\
      \hline
      \ion{N}{i} $^4S^0_{3/2}$         &  $\lambda$952, $\lambda\lambda\lambda$1134 & $16.37 \pm 0.06$  &$15.06 \pm 0.03$   & $< 13.7$                     \\
      \hline
      \ion{N}{v} $^2S_{1/2}$           &  $\lambda$1238, $\lambda$1242              & -                   &$13.56 \pm 0.03$   & -                          \\
      \hline
      \ion{O}{i} $^3P_{2}$             &  $\lambda$1039, $\lambda$1302              & $16.16 \pm 0.08$ (S)    & $ 15.53 \pm 0.37$ (S) & $ 15.68 \pm 0.06$ (S)                 \\
      \hline
      \ion{O}{i} $^3P_{1}$ (1s)        &  $\lambda$1304                             & $15.92 \pm 0.22$ (S)     & $ 14.17 \pm 0.18$ (S) & $ < 13.8$                  \\
      \hline
      \ion{O}{i} $^3P_{0}$ (2s)        &  $\lambda$1306                             & $14.38 \pm 0.04$     & $ 14.30 \pm 0.05$& $ < 13.8$                  \\
      \hline
      \ion{O}{vi} $^2S_{1/2}$          &  $\lambda$1031, $\lambda$1037              &                     & (S) &                           \\
      \hline
      \ion{Mg}{i} $^1S_0$              &  $\lambda$2026, $\lambda$2852              & $15.29 \pm 0.03$     & $ 14.01 \pm 0.13 $      &  $ 12.45 \pm 0.02 $                 \\
      \hline
      \ion{Mg}{ii} $^2S_{1/2}$          &  $\lambda$1239, $\lambda$1240, $\lambda$2796, $\lambda$2803& $16.28 \pm 0.02$&$15.03 \pm 0.22$& (S)        \\
      \hline
      \ion{Al}{ii} $^1S_0$              &  $\lambda$1670                             & $15.39 \pm 0.15$ (S)     &$16.42 \pm 0.08$ (S)    &$13.26 \pm 0.05$ (S)       \\
      \hline
      \ion{Al}{iii} $^2S_{1/2}$          &  $\lambda$1854, $\lambda$1862              & $13.21 \pm 0.02$    &$ 13.12 \pm 0.05$    &$12.55 \pm 0.11$       \\
      \hline
      \ion{Si}{ii} $^2P^{0}_{1/2}$       &$\lambda$1020, $\lambda$1190, $\lambda$1193, $\lambda$1260& $15.97 \pm 0.02$    &$15.78 \pm 0.03$    &$15.78 \pm 0.03$\\
      &$\lambda$1304, $\lambda$1526, $\lambda$1808&                    &         &                   \\
      \hline
      \ion{Si}{ii} $^2P^{0}_{3/2}$ (1s)   &  $\lambda$1264, $\lambda$1309, $\lambda$1533, $\lambda$1816 & $14.43 \pm 0.01$  &$15.19 \pm 0.23$   &$ < 13.5$      \\
  \hline
  \ion{Si}{iv} $^2S_{1/2}$          &  $\lambda$1393, $\lambda$1402              & -  &$14.16 \pm 0.04$      &  -                             \\
  \hline
  \ion{P}{i} $^4S^{0}_{3/2}$        & $\lambda$1674                              & $13.80 \pm 0.10$  &  $ < 13.7$            &  $ < 13.7$                 \\
  \hline
  \ion{P}{ii} $^3P_{0}$             &  $\lambda$1152                            & $13.61 \pm 0.04$  & $ 13.70 \pm 0.04$      & $ < 13.1$  \\
  \hline
  \ion{P}{v} $^2S_{1/2}$            &  $\lambda$1117                            & -                 &  $ 13.42 \pm 0.04$      & - \\
  \hline
  \ion{S}{ii} $^4S^{0}_{3/2}$       &$\lambda$1250, $\lambda$1253, $\lambda$1259& $15.57 \pm 0.02$  & $15.12 \pm 0.04  $ & $14.67\pm0.04$ \\
  \hline
  \ion{S}{vi} $^2S_{1/2}$           &  $\lambda$944                             & -                 & $ 14.21 \pm 0.03$ &   -  \\
  \hline
  \ion{Ar}{i} $^1S_{0}$             &  $\lambda$1048, $\lambda$1066,            & $14.60 \pm 0.04$  & $ 14.32 \pm 0.04 $  &  $13.59 \pm 0.07$\\
  \hline
  \ion{Ca}{i} $^1S_{0}$             &  $\lambda$4227                            & $12.14 \pm 0.02$  & -      &  $ < 11.1 $\\
  \hline
  \ion{Ca}{ii} $^2S_{1/2}$          &   $\lambda$3934, $\lambda$3969            & $15.21 \pm 0.03$  & $14.01 \pm 0.13$    & $ < 12.7 $       \\
  \hline
  \ion{Cr}{ii} $a^6S_{5/2}$           &$\lambda$2056, $\lambda$2062, $\lambda$2066& $13.89 \pm 0.02$  & $ 13.75 \pm 0.04$   & $ 13.16 \pm 0.07 $ \\
  \hline
  \ion{Fe}{ii} $a^6D_{9/2}$         &$\lambda$1081, $\lambda$1125, $\lambda$1143, $\lambda$1144& $15.42 \pm 0.03$  &$15.44 \pm 0.02$ & $14.03 \pm 0.11$\\
                                   &$\lambda$1608, $\lambda$1611, $\lambda$2249, $\lambda$2260&                   &       &\\
                                   &$\lambda$2344,$\lambda$2374, $\lambda$2382&                   &                     &\\
  \hline
  \ion{Fe}{ii} $a^6D_{7/2}$ (1s)    &$\lambda$1618,  $\lambda$1621, $\lambda$2389,  $\lambda$2396& $13.85 \pm 0.04$  & $ 14.09\pm 0.03$ & $ < 12.8$ \\
  \hline
  \ion{Fe}{ii} $a^6D_{5/2}$ (2s)    &$\lambda$1629,  $\lambda$1631, $\lambda$2328,  $\lambda$2405& $13.55 \pm 0.09$  & $ 13.87 \pm 0.04$&  $ < 13.4$\\
  \hline
  \ion{Fe}{ii} $a^6D_{3/2}$ (3s)    &$\lambda$1636,  $\lambda$2405, $\lambda$2407,  $\lambda$2411& $13.37 \pm 0.07$  & $ 13.55 \pm 0.07$ &  $ < 12.9$\\
  \hline
  \ion{Fe}{ii} $a^6D_{1/2}$ (4s)    &$\lambda$1639, $\lambda$2411, $\lambda$2414& $12.81 \pm 0.26$  & $13.16 \pm 0.12$  & $ < 12.8$ \\
  \hline
  \ion{Fe}{ii} $a^4F_{9/2}$ (5s)    &$\lambda$1566, $\lambda$1612, $\lambda$1637& $13.81 \pm 0.02$  & $13.83 \pm 0.02  $  & $ < 13.2$ \\
                                   &$\lambda$1658, $\lambda$1696, $\lambda$1702&   &     &       \\  
  \hline
  \ion{Fe}{ii} $a^4F_{7/2}$ (6s)    &$\lambda$1563, $\lambda$1580, $\lambda$1625& $12.82 \pm 0.18$   &$13.41 \pm 0.06$ &  $ < 12.4 $ \\
                                   &$\lambda$1659,  $\lambda$1712              &                   &                     &\\  
  \hline
  \ion{Fe}{ii} $a^4D_{7/2}$ (9s)    &  $\lambda$1635,  $\lambda$2740            &  $ < 12.8$        & $ 13.05 \pm 0.16      $   & $ < 12.8 $ \\
  \hline
  \ion{Ni}{ii} $^2D_{5/2}$          &$\lambda$1317, $\lambda$1370, $\lambda$1454& $14.22 \pm 0.03$  &$13.86 \pm 0.05$     & $13.80 \pm 0.06$\\
                                   &$\lambda$1709, $\lambda$1741, $\lambda$1751&                   &                     &\\
  \hline
  \ion{Ni}{ii} $^4F_{9/2}$ (2s)     &$\lambda$2166, $\lambda$2217, $\lambda$2223, $\lambda$2316& $13.89 \pm 0.04$  &$13.42 \pm 0.05$     & $ < 12.8 $ \\
  \hline
  \ion{Zn}{ii} $^2S_{1/2}$          &  $\lambda$2026,  $\lambda$2062            & $13.21 \pm 0.03$  &$12.86 \pm 0.07$    & $12.23 \pm 0.46$\\
  \hline
\end{tabular}

All values of the column densities are logarithmic  (in cm$^{-2}$). Upper limits are given at the $90\%$ confidence level.

$^1$(S) marks saturated transitions and unreliable column densities.}
\end{table*}

\section{Analysis}

In this section we analyze the absorption features of the host galaxy
detected in our X-shooter spectra. We report on the detected lines
(sect. 4.1), line-fitting procedure (sect. 4.2), low, excited and high
absorption features (sect. 4.3, 4.4 and 4.5, respectively), and the
hydrogen column density (sect. 4.6). Given the low S/N of the second
epoch spectrum, this is presented in sect. 4.7 only, to assess the
variability of some absorption features.

\subsection {Detected lines}

The host galaxy of GRB\,120327A is extremely rich in metal
lines. Metallic features are apparent from neutral ({\ion{N}{i}},
{\ion{O}{i}}, {\ion{Mg}{i}}, {\ion{P}{i}}, {\ion{Ar}{i}},
{\ion{Ca}{i}}), low-ionization ({\ion{C}{ii}}, {\ion{Mg}{ii}},
{\ion{Al}{ii}}, {\ion{Al}{iii}}, {\ion{Si}{ii}}, {\ion{P}{ii}},
{\ion{S}{ii}}, {\ion{Ca}{ii}}, {\ion{Cr}{ii}}, {\ion{Fe}{ii}},
{\ion{Ni}{ii}}, {\ion{Zn}{ii}}), and high-ionization ({\ion{C}{iv}},
{\ion{N}{v}}, {\ion{O}{vi}}, {\ion{Si}{iv}}, {\ion{P}{v}},
{\ion{S}{iv}}) species. In addition, strong absorption from the fine
structure levels of {\ion{C}{ii}}, {\ion{O}{i}}, {\ion{Si}{ii}},
{\ion{Fe}{ii}} and from the metastable levels of {\ion{Fe}{ii}} and
{\ion{Ni}{ii}} is identified, suggesting that the intense radiation
field from the GRB excites such features. Table~2 gives a summary of
all the absorption lines due to the host galaxy gas. The equivalent
widths of the most representative features are reported in Appendix
B. The spectral features were analyzed with FITLYMAN (Fontana \&
Ballester 1995). This program can simultaneously fit several
absorption lines, linking the redshifts, column densities and Doppler
parameters ($b$) if required.
The probed ISM of the host galaxy is resolved into three components
separated by $\sim 40$ km s$^{-1}$ from each other, which contribute
to the absorption system. The wealth of metal-line transitions allows
us to precisely determine the redshift of the GRB host galaxy,
$z=2.8145 \pm 0.0001$, setting the reference point to the central
component. This component features the strongest absorption from the
excited levels of the intervening gas with respect to the
corresponding ground state levels, implying that it is the closest
to the GRB (see next Section for details). This redshift is in
agreement with the values previously reported for GRB\,120327A (Perley
\& Tanvir 2012; Kr\"uhler et al. 2012; S\'anchez-Ramirez et al. 2012).

\subsection{Line-fitting procedure}

In order to dissect the GRB\,120327A environment into components, we
select a sample of absorption lines with the following properties: i)
they are in the X-shooter VIS arm (which has the highest resolution);
ii) they fall in a high signal-to-noise (S/N) wavelength region;
iii) they are far away from atmospheric lines and other spurious
features; iv) they are not saturated; v) they are produced by low
ionization transitions (high ionization levels are analyzed
separately, see Sect. 4.5). According to these criteria, we selected
the following lines to guide the identification of the different
components constituting the circumburst matter:
{\ion{Ni}{ii}~$\lambda\lambda$1741, 1751},
{\ion{Al}{iii}~$\lambda\lambda$1854, 1862},
{\ion{Ni}{ii}~$\lambda\lambda\lambda$2166, 2217, 2223} (the last multiplet
representing a metastable level of the {\ion{Ni}{ii}} ion).

A three-component model provides a good fit for this set of lines (see
Figs. 2 and 3), and this fit fixes the redshift of the three
components constituting the host galaxy absorption system. In the
following, we will identify component I as the bluest one, and
component III as the reddest one. The velocity separation between
component I and II (III) is $41$ ($76$) km~s$^{-1}$. We stress that
the number of identified components in a GRB environment, apart from
reflecting the complexity of the specific host galaxy ISM, depends on
the resolution of the instrument used. The nominal resolution of the
X-shooter VIS arm is $R=8800$. However, during our observations, the
seeing was down at $0.5 \arcsec$ for most of the integration time, and
this increased the actual resolution. To estimate the true resolution
of our observations, we used the width of the telluric lines, finding
that $R=12000$ in the VIS and $R=7000$ in the UVB and NIR arms
represent more reliable values for our data. A resolution of $R=12000$
translates into an instrumental $b$ parameter of $v_R=15$ km~s$^{-1}$
in velocity space. This means that we can not resolve line separations
much smaller than $v_R$.

The redshifts of our three components were fixed in order to fit the
other species present in the spectrum with the same model (high
ionization lines are analyzed in a different way, see Sect. 4.3). The
Doppler parameter $b$ was fixed between ions and levels of the same
atom and was allowed to vary between different atoms.  $b$ is in
  the range $15-40$ km $s^{-1}$, $5-30$ km $s^{-1}$ and $25-100$ km $s^{-1}$
  for components I, II and III respectively.  In some cases the fit
returned a $b$ value for component II smaller than the spectral
resolution. To assess the robustness of the column density values for
these specific lines, we verified that the column density obtained
fixing $b$ to $15$ km s$^{-1}$ was within $1 \sigma$ with respect to
that obtained leaving $b$ free to vary. Table 2 summarizes all the
species and the corresponding transitions identified in our
spectrum. The reported uncertainties are the formal $1 \sigma$ errors
provided by FITLYMAN, while the upper limits are at the 90\%
confidence level. The next sub-sections describe the detected lines in
detail.

\begin{figure*}
\centering
\includegraphics[angle=-0,width=18cm]{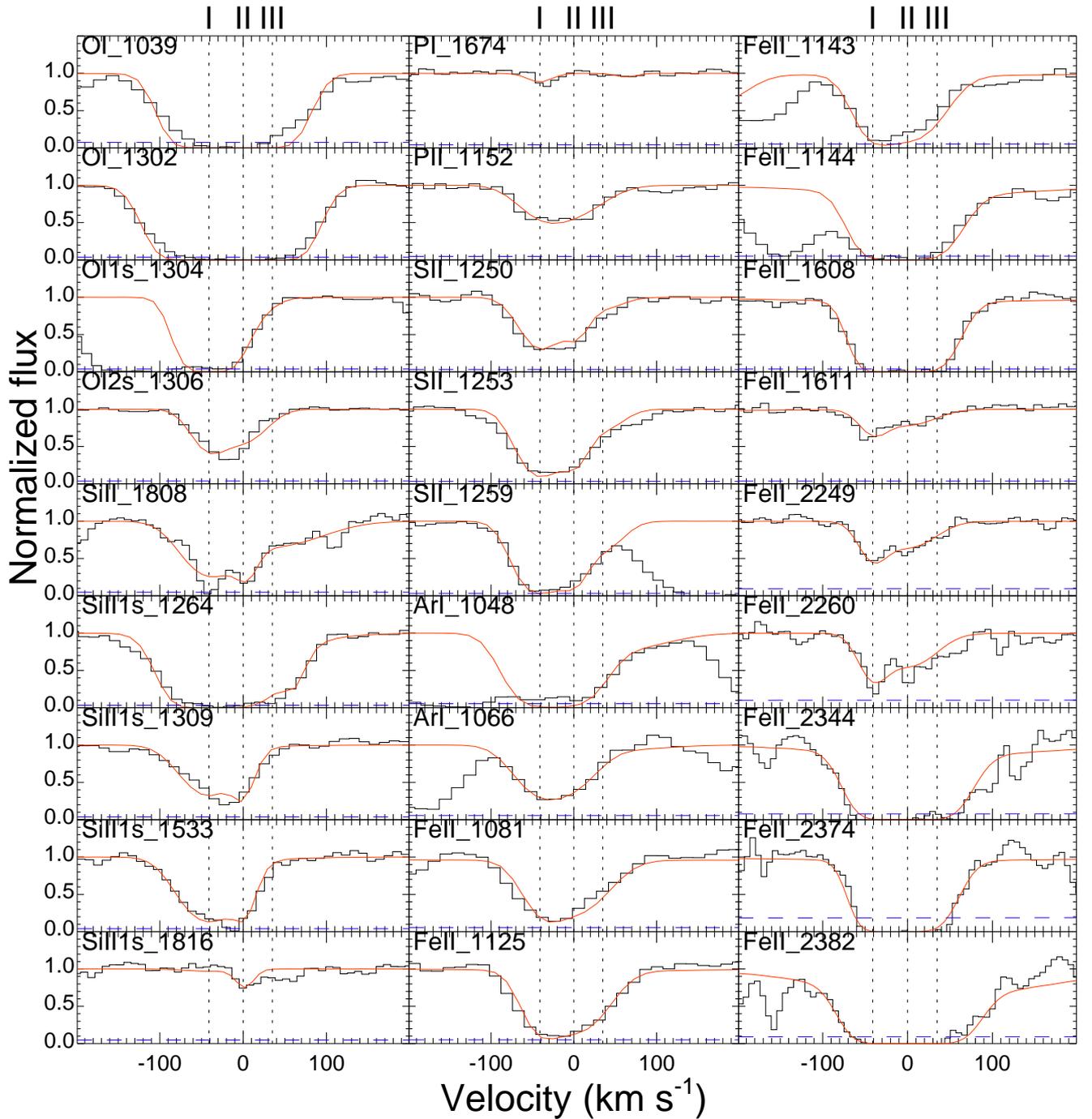}
\caption{The GRB\,120327A absorption features. Solid, red lines
  represent the best-fit model to the three Voigt components. Vertical
  dashed lines identify the component velocities. Blue, dashed lines
  display the error spectrum. The zero point has been arbitrarily
  placed at the redshift of the second component ($z=2.8145$).  }
\label{spe1}
\end{figure*}

\begin{figure*}
\centering
\includegraphics[angle=-0,width=18cm]{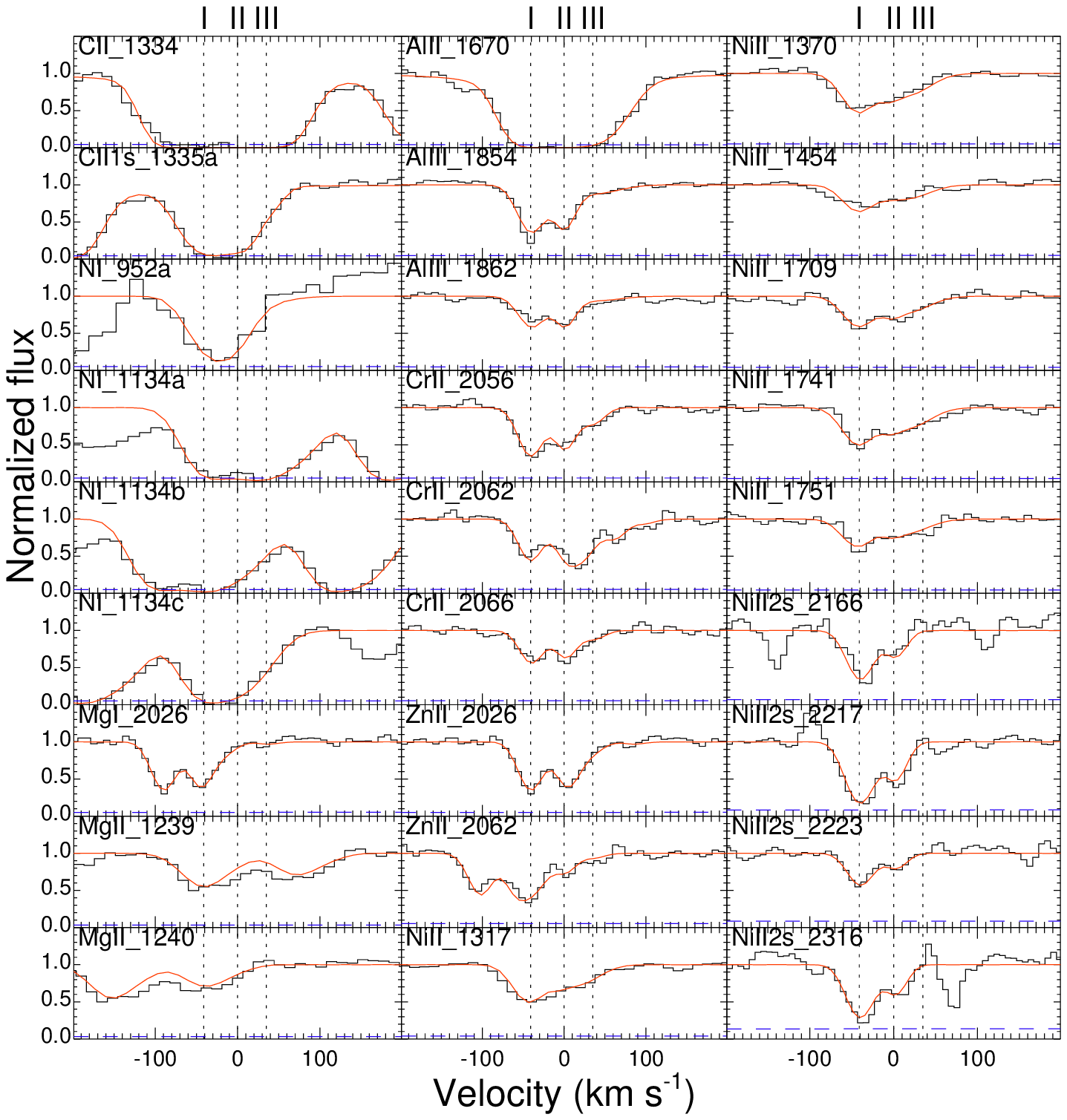}
\caption{The GRB\,120327A absorption features. Solid, red lines
  represent the best-fit model to the three Voigt components. Vertical
  dashed lines identify the component velocities. Blue, dashed lines
  display the error spectrum. The zero point has been arbitrarily
  placed at the redshift of the second component ($z=2.8145$). The
  absorption on the left of the {\ion{Mg}{i}$~\lambda$2026} line is
  due to {\ion{Zn}{ii}~$\lambda$2026}.}
\label{spe1}
\end{figure*}

\subsection {Low ionization lines}

We briefly describe here all transitions due to neutral, singly- or
doubly-ionized species. The {\ion{C}{ii}~$\lambda$1334},
{\ion{O}{i}$~\lambda$1039}, {\ion{O}{i}$~\lambda$1302} and
{\ion{Al}{ii}~$\lambda$1670} transitions are heavily saturated, thus
the computed column densities of these species are unreliable or
should be regarded with caution (they are marked with an ``S'' in
Tab. 2). In addition, some level of saturation might also be present
in component I of the {\ion{Si}{ii}~$\lambda$1808} line, the only one
available to determine the \ion{Si}{ii} column density. We note that
the {\ion{Mg}{ii}~$\lambda$2796} and {\ion{Mg}{ii}~$\lambda$2803}
lines are heavily saturated too, but we can determine the
{\ion{Mg}{ii}} column density using the weak
{\ion{Mg}{ii}~$\lambda\lambda$1239, 1240} doublet.  Figures 2, 3 and 4
show all the transitions used to determine the column densities, and
the corresponding best-fit models. In particular, Figures 2 and 3
display transitions observed in the UVB and VIS arms, while Fig. 4
shows NIR arm transitions, where the resolution is lower and the
components are not well separated. In each figure the three components
identified by our analysis are marked by vertical dashed lines.

All singly and doubly ionized features, with the excpetion of
\ion{P}{ii} and \ion{Ca}{ii}, are detected in each of the three
components. It is evident, however, that component III is the least
populated of the three, while components I and II show similar column
densities, with the exception of \ion{Ca}{ii}, which is stronger in
component I.

The neutral species ({\ion{N}{i}}, {\ion{O}{i}}, {\ion{Mg}{i}},
{\ion{P}{i}}, {\ion{Ar}{i}}, {\ion{Ca}{i}}) are considerably weaker in
component III (if detected). In addition, \ion{P}{i} and \ion{Ca}{i}
show absorption only in component I, and the \ion{N}{i} and
{\ion{Mg}{i}} column densities of component I are considerably higher
than that of component II. The strong saturation of the {\ion{O}{i}}
features prevents similar considerations for this species.

\begin{figure}
\centering
\includegraphics[angle=-0,width=9cm]{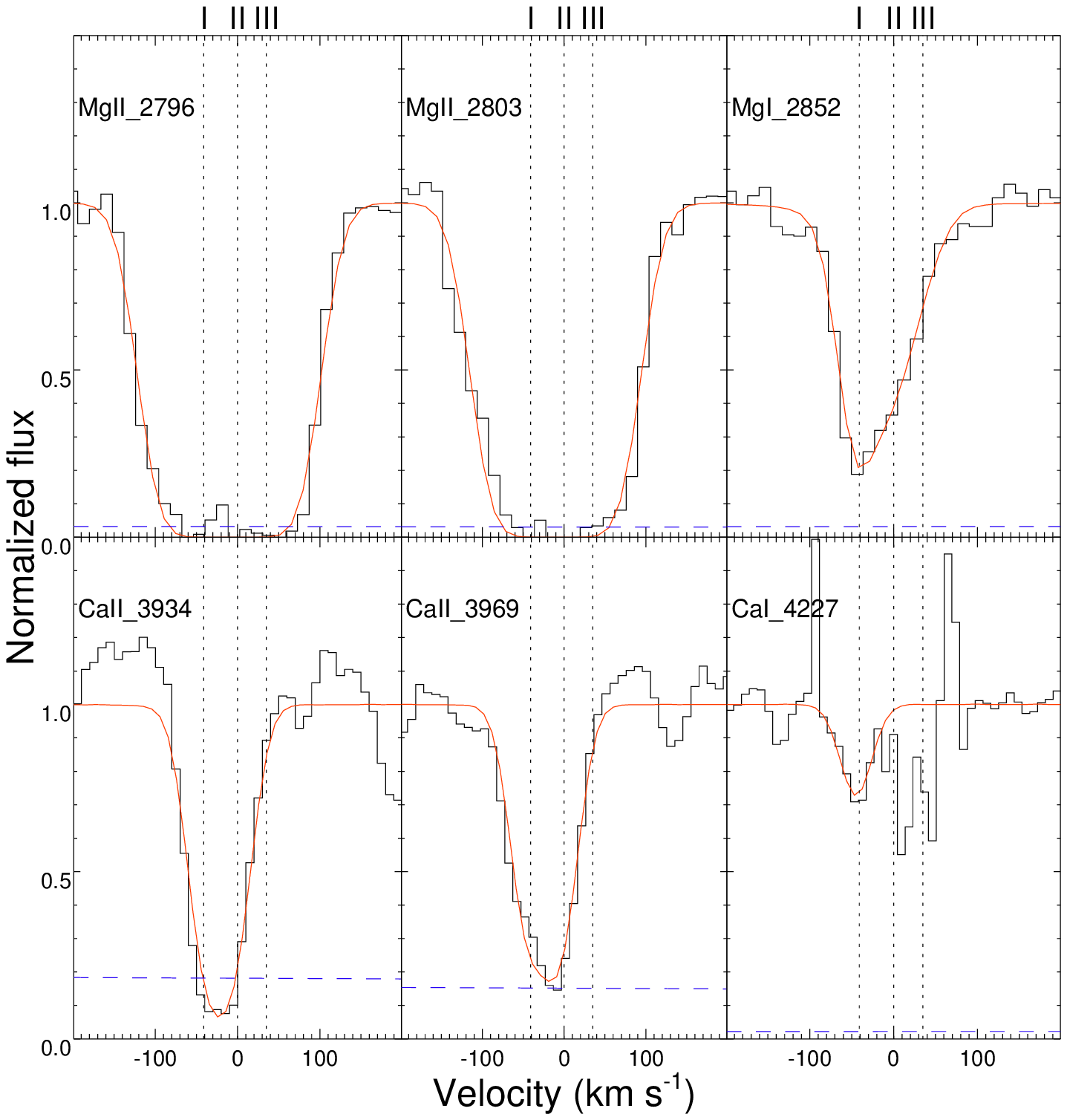}
\caption{The GRB\,120327A absorption features detected in the
  X-shooter NIR arm. Solid, red lines represent the best-fit model to
  the three Voigt components. Blue, dashed lines display the error
  spectrum. Vertical dashed lines identify the component
  velocities. Note that the resolution in this arm is lower than in
  the VIS one, so that single components do not appear well
  separated.}
\label{spe1}
\end{figure}

\subsection {Excited levels}

The level structure of an atom or ion is characterized by a principal
quantum number $n$, which defines the atomic level, and by the
spin-orbit coupling (described by the quantum number $j$), which
splits these levels into fine structure sublevels. In GRB absorption
spectra, several excited features are detected at the GRB redshift,
due to the population of both $n>1$ and/or $n=1$ fine structure
levels. GRB\,120327A is not an exception, featuring excited levels of
{\ion{C}{ii}}, {\ion{O}{i}}, {\ion{Si}{ii}, {\ion{Fe}{ii}} and
{\ion{Ni}{ii}}. In particular, {\ion{C}{ii}} and {\ion{Si}{ii}} show
fine structure lines produced by the first excited $n=1$ level, while
{\ion{O}{i}} shows fine structure lines from the first and second
excited $n=1$ levels (Figs. 2 and 3). {\ion{Ni}{ii}} features instead
absorption from the ground level of the $n=2$ quantum state (Fig. 3).
Finally, {\ion{Fe}{ii}} shows a plethora of absorption lines due to at
least seven excited levels belonging to the $n=1$, $n=2$ and $n=3$
quantum states (Fig. 5). 

The excited transitions displayed in Figs. 2, 3 and 5 are marked with a
number that indicates the level above the $n=1$ fundamental state to
adopt a compact notation (see Table 2 for the standard definition
based on their quantum numbers). For example, {\ion{C}{ii}}1s is one
level above the fundamental state and so on. {\ion{Ni}{ii}}2s
represents the ground state level of the $n=2$ quantum state for the
singly-ionized Nickel, while \ion{Fe}{ii}5s and \ion{Fe}{ii}9s are the
ground state levels of the $n=2$ and $n=3$ quantum states for the
singly-ionized Iron, respectively.

As for the low ionization lines, saturation problems arise when one
tries to fit the \ion{C}{ii} and \ion{O}{i} excited lines. In
particular, the {\ion{O}{i}1s$~\lambda$1304} fine structure line is
heavily blended with the {\ion{Si}{ii}~$\lambda$1304} feature. The
computation of the corresponding column density has been attempted by
simultaneously fitting the {\ion{O}{i}1s~$\lambda$1304} and
{\ion{Si}{ii}~$\lambda$1304} lines, fixing the column densities of the
latter to that evaluated using non-saturated transitions. Despite
this, we cannot exclude strong saturation effects, so the computed
column densities of the \ion{C}{ii} and \ion{O}{i} fine structure
levels should be considered with caution.

Excited levels are not present in component III, just like neutral
species. However, while the latter tend to populate more heavily
component I (see Table 2 and previous sub-section), the former are
more prominent in component II, when compared to their corresponding
ground state level (with the exception of {\ion{Ni}{ii}}2s and
\ion{Fe}{ii}5s).

\subsection {High-ionization lines}

The high-ionization lines detected in the GRB\,120327A spectrum belong
to the following species: {\ion{C}{iv}}, {\ion{N}{v}}, {\ion{O}{vi}},
{\ion{Si}{iv}}, {\ion{P}{v}} and {\ion{S}{vi}}. We prefer to adopt a
different approach while fitting high ionization lines because i) some
of them appear to be saturated (in particular the {\ion{C}{iv}} and
{\ion{O}{vi}} features); ii) most of them fall in the X-shooter UVB
arm, where resolution is lower; iii) a physical separation between
low- and high- ionization species, and thus different line profiles,
is not unexpected, because they may trace different ISM regions
because of their different ionization potentials. For these reasons,
we try to fit the high-ionization lines with a single Voigt profile
with linked central wavelength. There is a good alignment between all
the absorption lines, with the exception of the {\ion{N}{v}}
features. For this doublet, we have to shift the central wavelength in
order to obtain a satisfactory fit.

Fig. 6 shows the high-ionization lines and the best-fit model. In the
last panel the {\ion{Al}{iii}~$\lambda$1854} low-ionization line is
shown for comparison. The vertical dotted lines mark the central
velocity derived for the {\ion{N}{v}} features (right lines) and that
for all the other high-ionization species (left lines). {\ion{N}{v}}
appears perfectly lined-up with component II of the low-ionization
species, while the central wavelength of the other high-ions falls
exactly halfway between component I and II.

\begin{figure*}
\centering
\includegraphics[angle=-0,width=18cm]{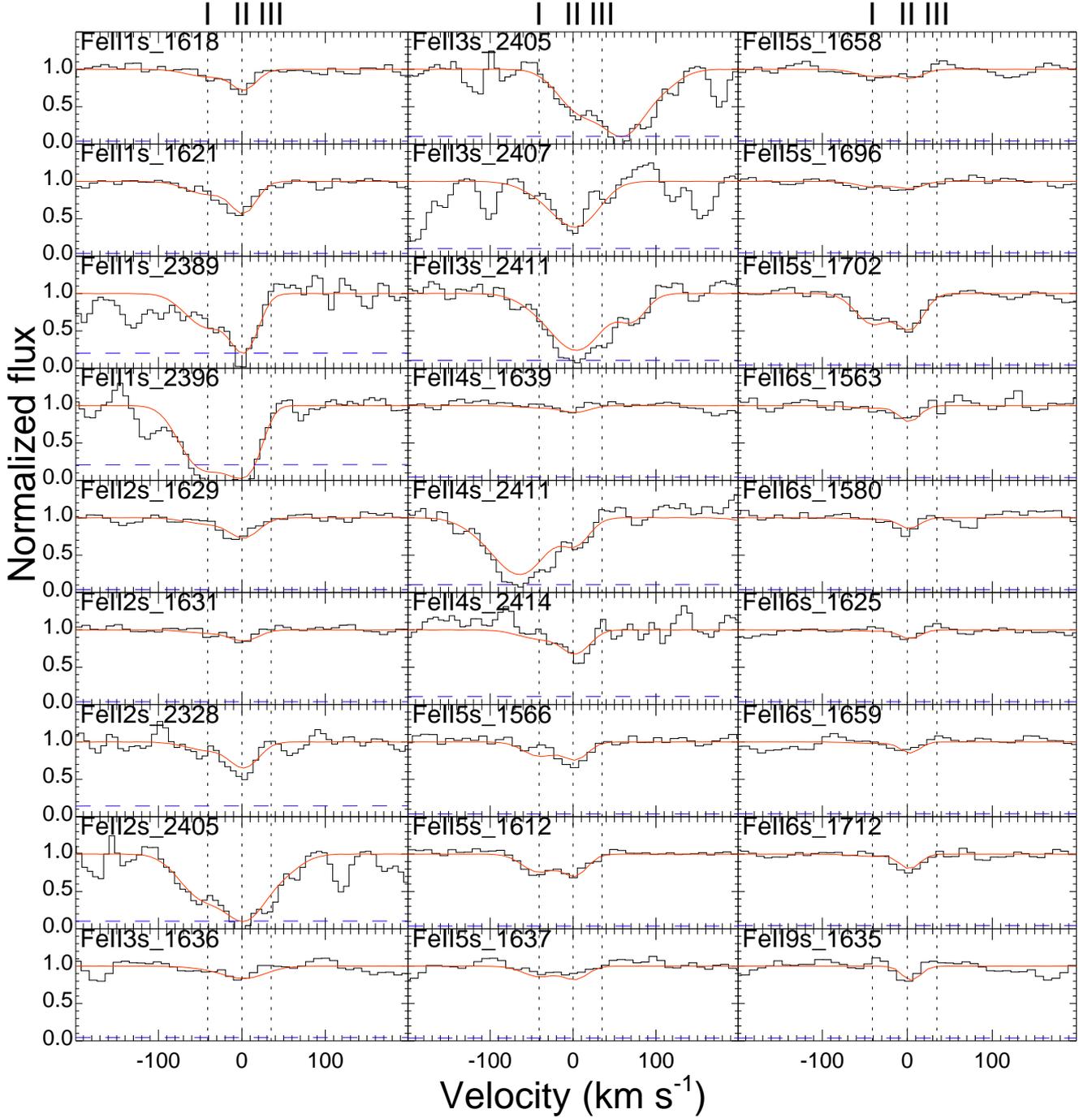}
\caption{The {\ion{Fe}{ii}} excited absorption features in the
  GRB\,120327A afterglow spectrum. Solid, red lines represent the
  best-fit model with three Voigt components. Blue, dashed lines
  display the error spectrum. Vertical dashed lines identify the
  component velocities. The zero point has been arbitrarily placed at
  the redshift of the second component ($z = 2.8145$).}
\label{spe1}
\end{figure*}

\begin{figure}
\centering
\includegraphics[angle=-0,width=9cm]{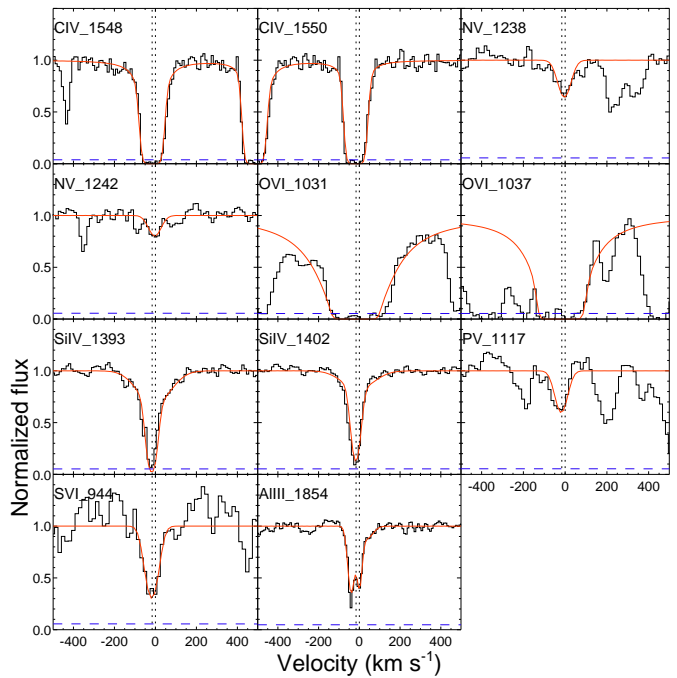}
\caption{The GRB\,120327A high ionization absorption features. Solid,
  red lines represent the best-fit model with a single Voigt
  component. Blue, dashed lines display the error spectrum.  Vertical
  dashed lines identify the component velocities. In particular, the
  rightmost line indicates the center of the {\ion{N}{v}} profile, while
  the left one the center of all the other lines.}
\label{spe1}
\end{figure}

\begin{figure}
\centering
\includegraphics[angle=-0,width=9cm,height=6.5cm]{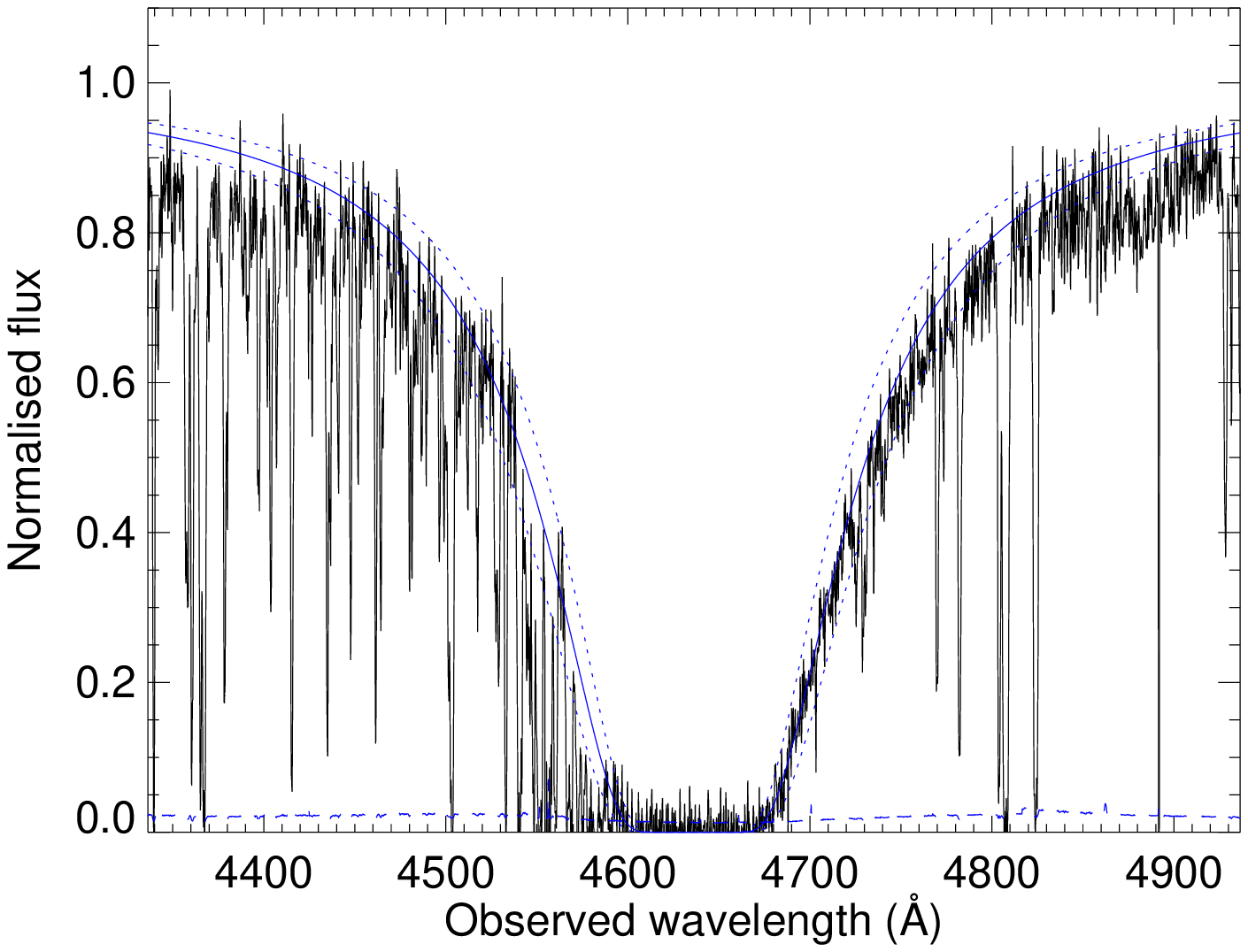}
\includegraphics[angle=-0,width=8.cm]{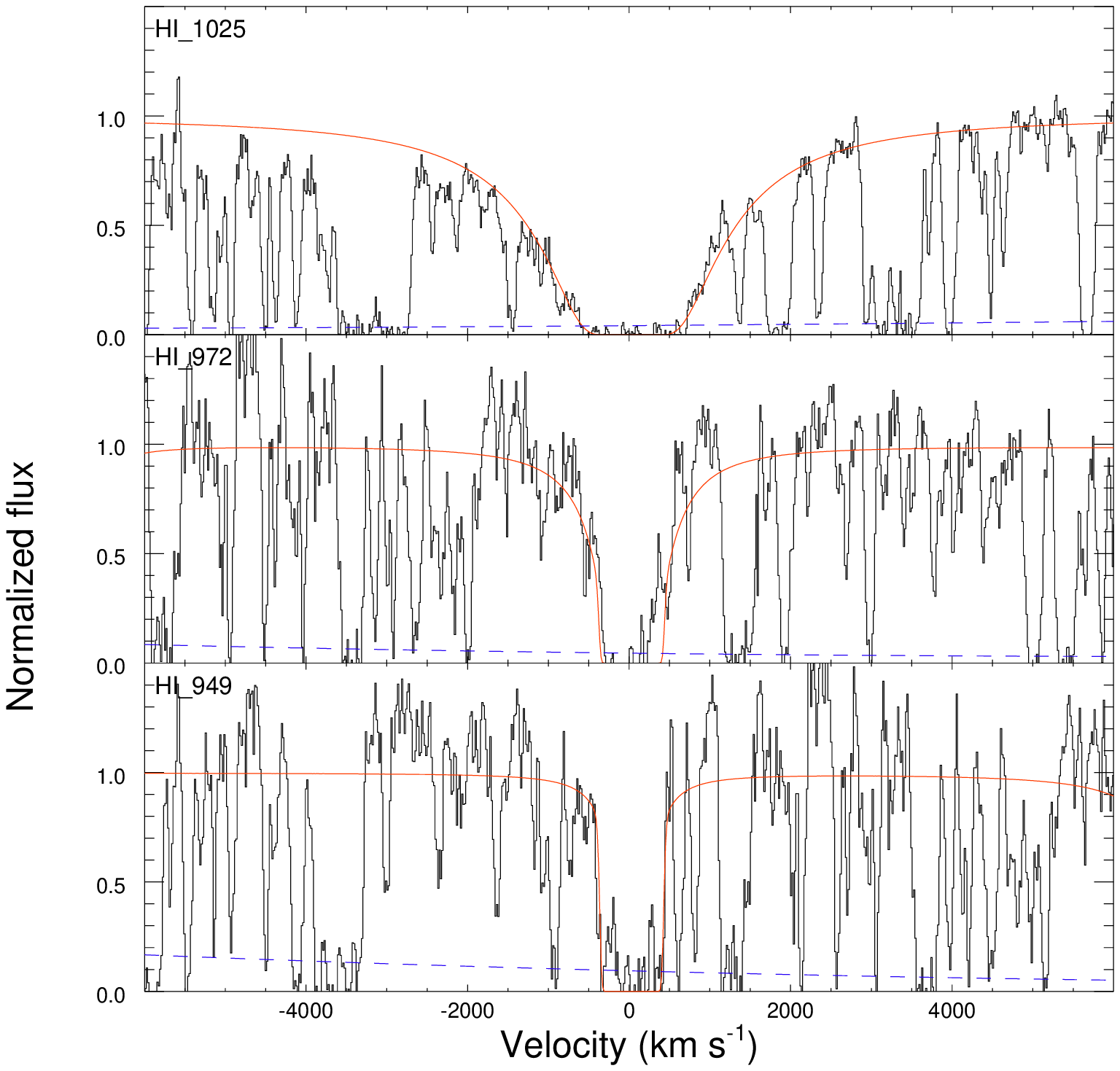}
\caption{Top panel: the Ly$\alpha$ absorption feature at the
  GRB\,120327A redshift. The solid, blue line represents the best-fit
  model to the absorption profile; the dotted, blue lines enclose the
  $1 \sigma$ uncertainty. Bottom panel: the Ly$\beta$, Ly$\gamma$ and
  Ly$\delta$ features (from top to bottom) and the corresponding fit
  result (solid, red lines). Dashed blue lines displays the error
  spectrum.}
\label{spe1}
\end{figure}

\subsection{Hydrogen column density}

The GRB\,120327A redshift is high enough to allow several hydrogen
lines belonging to the Lyman series to enter the X-shooter spectral
window. Ly$\alpha$ and Ly$\beta$ are clearly seen at $\sim 4650$ and
$3900$ \AA, respectively. Given the ADC problem in the X-shooter
observation of GRB\,120327A, the normalization around the Ly$\alpha$
is not easy, since it is difficult to separate the contribution of the
continuum to that of the Ly$\alpha$ wings. On the other hand, the
fitting of Lyman lines bluewards of the Ly$\alpha$ must be performed
with caution, since the Ly$\alpha$ forest affects relevant
contamination to the line profiles. We thus used two different
approaches to determine the hydrogen column density, and checked the
consistency of the corresponding results. To fit the Ly$\alpha$, we
first extracted the spectrum from the two dimensional one using a
standard response curve for the fluxing. We then dereddened it for
Galactic extinction and normalised with a power-law $F_\lambda \propto
\lambda^{-0.3}$. A Ly$\alpha$ feature was then fitted to this
normalized spectrum. We note that this spectral index is not
consistent with that deduced from photometric measures (Sect. 5.2) or
from the GTC spectrum (Sect. 5.5), due to the ADC problem. Figure 7
(top panel) shows the result of the fit with $\log (N_{\rm H,1}/{\rm
  cm}^{-2}) = 22.00 \pm 0.10$ cm$^{-2}$. Dotted lines delimit the $1
\sigma$ uncertainty.
To study other Lyman features, we normalized the spectrum locally,
using the few line-free regions to determine the continuum
contribution, and simultaneously fit the Ly$\beta$, Ly$\gamma$ and
Ly$\delta$, fixing the redshift at that of the central metallic
component (i.e., $z=2.8145$). Bluer Lyman features are present (up to
at least the Ly$\theta$) but the S/N in the corresponding regions is
too low to enable a good fit. Figure 7 (bottom panel) shows the
result of the fit with $\log (N_{\rm H,2}/{\rm cm}^{-2}) = 22.05 \pm
0.20$. The two results are consistent, and the best value for the
hydrogen column density is obtained with a weighted average of them,
yielding $\log (N_{\rm H}/{\rm cm}^{-2}) = 22.01 \pm 0.09$. All the
Lyman features are damped and cannot be separated into components, as
we did for the metallic lines. The line fits are virtually insensitive
to the adopted $b$ parameter.

\subsection{Second-epoch spectroscopy }

A second X-shooter spectrum was acquired $\approx 30$~hr after the
GRB. Unfortunately, the S/N for this observation is extremely poor
(see Table 1), since the afterglow was considerably fainter. However,
we could compare the strongest {\ion{Fe}{ii}} ground state transitions
between the two epochs. Columns 2 and 3 of Tab. 3 indicate whether the
corresponding transition in column 1 is blended with {\ion{Fe}{ii}}
fine structure lines or contaminated by atmospheric lines. Columns 4
and 5 report the rest-frame equivalent widths (EWs) of the transitions
in the first and second epoch, respectively. Column 6 and 7 give the
variation and significance between the two observations, respectively.
All lines not contaminated by telluric features show no variation at
more than the $3 \sigma$ level. For the {\ion{Fe}{ii}}~$\lambda$2344,
{\ion{Fe}{ii}}~$\lambda$2374 and {\ion{Fe}{ii}}~$\lambda$2382 features
the reported variation is possibly induced by a strong contamination
by sky lines. In fact, after applying the proper telluric correction,
the discrepancy between the two epochs falls below the $3 \sigma$
level. Thus, at the $3\sigma$ confidence level, the {\ion{Fe}{ii}}
ground state is not varying. We applied the same procedure to the
{\ion{Fe}{ii}} fine structure line with the strongest oscillator
strength, i.e., the {\ion{Fe}{ii1s}}~$\lambda$2396 feature. In this
case, the EW in the second epoch is compatible with zero, and the
significance of the variation with respect to the first epoch is more
than $3 \sigma$ (Tab. 3).  A simple inspection of the two spectra
reveals that all the excited {\ion{Fe}{ii}} lines disappeared in the
second observation. However, the $3 \sigma$ confidence level is
reached for the {\ion{Fe}{ii1s}}~$\lambda$2396 feature only. Using
FITLYMAN, we could determine a $90\%$ upper limit to its column
density of $\log (N/{\rm cm}^{-2}) < 13.5$.

\begin{table*}[ht]
\centering
\caption{\bf Comparison between the observed {\ion{Fe}{ii}} equivalent widths in the first and second X-shooter observation}
{\footnotesize
\smallskip
\begin{tabular}{|l|cccccc|}
\hline
        
Feature	  &	Blended?&	Atm?&	        EW1 (\AA)&			EW2 (\AA)	&	Variation (\AA)&		Significance ($\sigma$)\\
\hline	
\ion{Fe}{ii} $\lambda$1608&		No&	        No&		$3.11\pm0.04$&		$3.3\pm0.9$&	$+0.19\pm0.94$&	$+0.2$\\
\ion{Fe}{ii} $\lambda$2250&		No&		No&		$1.22\pm0.05$&		$1.54\pm0.17$&	$+0.32\pm0.22$&	$+1.5$\\
\ion{Fe}{ii} $\lambda$2261&		No&		No&     	$1.58\pm0.06$&		$1.81\pm0.23$&	$+0.23\pm0.29$&	$+0.8$\\
\ion{Fe}{ii} $\lambda$2344&		Yes&		Yes&		$5.41\pm0.11$&		$3.4\pm0.2$&	$-2.01\pm0.31$&	$-6.5$\\
\ion{Fe}{ii} $\lambda$2374&		No&		Yes&		$4.11\pm0.06$&		$3.05\pm0.18$&	$-1.06\pm0.24$&	$-4.4$\\
\ion{Fe}{ii} $\lambda$2382&		Yes&		Yes&		$7.13\pm0.09$&		$4.08\pm0.19$&	$-3.05\pm0.28$&	$-10.9$\\
\ion{Fe}{ii} $\lambda$2587&		No&		No&		$5.81\pm0.29$&		$3.80\pm0.47$&	$-2.01\pm0.76$&	$-2.6$\\
\ion{Fe}{ii} $\lambda$2344&		Yes&		Corrected&	$4.85\pm0.16$&		$3.40\pm0.45$&	$-1.44\pm0.61$&	$-2.4$\\
\ion{Fe}{ii} $\lambda$2374&		No&		Corrected&	$3.99\pm0.12$&		$2.48\pm0.39$&	$-1.51\pm0.51$&	$-2.9$\\
\ion{Fe}{ii} $\lambda$2382&		Yes&		Corrected&	$4.71\pm0.16$&		$3.13\pm0.45$&	$-1.58\pm0.61$&	$-2.6$\\
\ion{Fe}{ii} $\lambda$2396*&		Yes&		Corrected&	$1.80\pm0.12$&		$0.24\pm0.33$&	$-1.56\pm0.45$&	$-3.5$\\

\hline
\end{tabular}
}
\end{table*}

\section{Results}

In this section we extract the physical information from the column
densities estimated through our analysis. In Sect. 5.1 we derive the
metallicity for the host galaxy of GRB\,120327A; in Sect 5.2 we
compute the absorber-GRB distance comparing the data with
photo-excitation codes; the characterization of the GRB\,120327A host
is presented in Sect. 5.3 through the analysis of the dust depletion
patterns and chemical evolution models; in Sect. 5.4 we estimate the
H$_2$ molecule column density and the corresponding molecular
fraction; finally, in Sect. 5.5 we derive the extinction curve shape
for the GRB\,120327A afterglow.

\subsection{Abundances}

We compute here the metal abundances (relative to hydrogen)
of the GRB\,120327A host, both in absolute terms and compared
with solar values, using the data reported in Table 2. The wide
spectral coverage of X-shooter allows this computation for several
elements. We proceeded as follows. First, we summed the column
densities of the transitions belonging to the same atom (different
ionization and excitation states), in order to evaluate the atomic
column densities for each of the three components (we consider neutral
and low-ionization state only); second, since the Lyman features
cannot be resolved into components, we summed up the atomic column
densities of the three components to obtain the total column density
for each species. We note that our (non-saturated) column densities
are always dominated by a single ionic species, with the only
exception of Phosphorus, where {\ion{P}{i}} and {\ion{P}{ii}} give a
comparable contribution to the total column density.

The results are listed in Table 4. Columns 2, 3 and 4 report the total
column densities for each element in components I, II and III,
respectively. The total column density (obtained summing up the
previous three columns) is given in column 5. Column 6 reports the
total abundances relative to hydrogen in the GRB\,120327A host, while
column 7 reports the element abundances in the solar photosphere as
published by Asplund et al. (2009). Finally, column 9 gives the
abundances of the GRB\,120327A host relative to the solar values. We
derived metallicity values with respect to the solar ones between
$1.9\times10^{-2}$ and $7.8\times10^{-2}$ for most of the
elements. Exceptions to this behavior are represented by \ion{C}{} and
\ion{Al}{}, for which Solar or supersolar values are instead
obtained. This deviation is most likely due to the fact that the
derived column densities of these elements suffer from strong
saturation problems (as that of \ion{O}{}), so the corresponding
abundances are not reliable.


\begin{table*}[ht]
\centering
\caption{\bf Metallicities}
{\footnotesize
\smallskip
\begin{tabular}{|l|ccc|c|c|c|c|}
\hline
           &    \multicolumn{4}{c|}{$\log N_{\rm X} /{\rm cm}^{-2}$}        &$\log N_{\rm X}$/$N_{\rm H}$ & $(\log N_{\rm X}$/$N_{\rm H})^a_{\odot}$  &$[{\rm X}/{\rm H}]$     \\
\hline
Element X  & Component I    & Component II       & Component III      & Total          &   &   &\\
\hline
C$^{(S)}$& $16.22\pm0.26$ & $18.65\pm0.15$ & $14.99\pm0.20$ & $18.65\pm0.15$ & $-3.36\pm0.17$ & $-3.57\pm0.05$ &$0.21\pm0.18$    \\
N      & $16.37\pm0.06$ & $15.06\pm0.03$ & $-$            & $16.39\pm0.06$ & $-5.62\pm0.10$ & $-4.17\pm0.05$ &$-1.45\pm0.11$   \\
O$^{(S)}$& $16.69\pm0.08$ & $15.57\pm0.33$ & $-$            & $16.72\pm0.08$ & $-5.29\pm0.12$ & $-3.31\pm0.05$ &$-1.98\pm0.13$    \\
Mg     & $16.32\pm0.02$ & $15.07\pm0.20$ & $12.45\pm0.02$ & $16.34\pm0.02$ & $-5.67\pm0.09$ & $-4.40\pm0.04$ &$-1.27\pm0.10$   \\
Al$^{(S)}$& $15.39\pm0.15$ & $16.42\pm0.08$ & $13.34\pm0.05$ & $16.46\pm0.07$ & $-5.55\pm0.11$ & $-5.55\pm0.03$ &$0.00\pm0.11$   \\
Si     & $15.98\pm0.02$ & $15.88\pm0.05$ & $15.78\pm0.03$ & $16.36\pm0.03$ & $-5.65\pm0.09$ & $-4.49\pm0.03$ &$-1.16\pm0.09$   \\
P      & $14.02\pm0.06$ & $13.70\pm0.04$ & $-$            & $14.19\pm0.04$ & $-7.82\pm0.10$ & $-6.59\pm0.03$ &$-1.23\pm0.10$   \\
S      & $15.57\pm0.02$ & $15.12\pm0.04$ & $14.67\pm0.04$ & $15.74\pm0.02$ & $-6.27\pm0.09$ & $-4.88\pm0.03$ &$-1.39\pm0.09$   \\
Ar     & $14.60\pm0.04$ & $14.32\pm0.04$ & $13.59\pm0.07$ & $14.81\pm0.03$ & $-7.20\pm0.09$ & $-5.60\pm0.13$ &$-1.60\pm0.15$   \\
Ca     & $15.21\pm0.03$ & $14.01\pm0.13$ & $-$            & $15.24\pm0.03$ & $-6.77\pm0.09$ & $-5.66\pm0.04$ &$-1.11\pm0.10$   \\
Cr     & $13.89\pm0.02$ & $13.75\pm0.04$ & $13.16\pm0.07$ & $14.17\pm0.02$ & $-7.84\pm0.09$ & $-6.36\pm0.04$ &$-1.48\pm0.09$   \\
Fe     & $15.45\pm0.03$ & $15.49\pm0.02$ & $14.03\pm0.11$ & $15.78\pm0.02$ & $-6.23\pm0.09$ & $-4.50\pm0.04$ &$-1.73\pm0.10$   \\
Ni     & $14.39\pm0.02$ & $13.99\pm0.04$ & $13.80\pm0.06$ & $14.61\pm0.02$ & $-7.41\pm0.09$ & $-5.78\pm0.04$ &$-1.63\pm0.09$   \\
Zn     & $13.21\pm0.03$ & $12.86\pm0.07$ & $12.23\pm0.46$ & $13.40\pm0.04$ & $-8.61\pm0.10$ & $-7.44\pm0.05$ &$-1.17\pm0.11$   \\

\hline
\end{tabular}
}

$a$: Solar abundances taken from photospheric values in Asplund et al. (2009).
$(S)$: Abundances derived from saturated transitions, thus not reliable. 
\end{table*}

\subsection{Distance between the GRB and the absorbers}

As shown in the previous section, a wealth of excited lines from
{\ion{C}{ii}}, {\ion{O}{i}}, {\ion{Si}{ii}}, {\ion{Ni}{ii}}, and
especially from {\ion{Fe}{ii}} are detected. 
The strong variability in the column density of the {\ion{Fe}{ii}} and
{\ion{Ni}{ii}} excited levels observed in GRB\,060418 (Vreeswijk et
al. 2007), GRB\,080319B (D'Elia et al. 2009a), and GRB\,080310 (De Cia
et al. 2012) ruled out collisional processes and direct infrared
pumping as responsible for the excitation. This leaves us with the
indirect UV pumping as responsible for the excited level population
(Prochaska et al. 2006). A collisional origin of the excitation can be
ruled out even if multi-epoch spectroscopic data are missing. In fact,
collisions populate higher energy levels less than lower energy
ones. For instance, the spectra of GRB\,080330 (D'Elia et al. 2009b),
GRB\,050730 (Ledoux et al. 2009) and GRB\,090926A (D'Elia et al. 2010)
exhibit an \ion{Fe}{ii} $a^4F_{9/2}$ excited state column density that
is higher than several fine structure levels of the \ion{Fe}{ii}
$a^6D$ ground state. This means that radiative processes are also at
work for these GRBs.

For GRB\,120327A we have multi-epoch spectroscopy, with the second
observation following the first one by about one day. Despite the low
S/N of the second epoch spectrum, in section 4.7 we reported that the
\ion{Fe}{ii}~$\lambda$2396 fine structure line disappeared in this
time interval with a high confidence level (Table 3). This proves that
UV pumping, rather than collisions, excited the atomic species in the
GRB\,120327A host galaxy gas. Anyway, even if the second epoch
spectroscopy were not available, the higher column density of the
\ion{Fe}{ii} $a^4F_{9/2}$ excited level with respect to lower-excited
\ion{Fe}{ii} states would be sufficient to assess that radiative
processes are at work.


It is straightforward to note that the excited levels are directly
influenced by the flux impacting the intervening gas. This means that
the distance between the GRB and the absorbers can be estimated from
the column densities of the excited states and their ratio. To
estimate this distance, we compare the observed column densities to
those predicted by a photo-excitation code for the time when the
spectroscopic observations were acquired. The photo-excitation code is
that used by Vreeswijk et al. (2007) and D'Elia et
al. (2009)\footnote{Taking into account the corrigendum reported in
  Vreeswijk et al (2011).}, to which we refer the reader for more
details. Basically, it solves the detailed balance equation in a
time-dependent way for a set of transitions involving the levels of a
given species (e.g. \ion{Fe}{ii} and \ion{Si}{ii}).  The equation
depends on the flux level experienced by the absorbing gas. This flux
is, of course, a function of the distance $d$ of the gas from the GRB
explosion site, which is a free parameter of the computation and the
quantity we want to calculate. The other free parameters are the
initial column densities of the levels involved and the Doppler
parameter of the gas itself. Since our X-shooter spectrum suffers from
ADC problems, we are not able to estimate a reliable afterglow
spectral index from our dataset. Thus, we used the photometry reported
by Sudilovsky et al. (2013), who observed the afterglow simultaneously
in the $g'r'i'z'JHK$ bands with GROND (Greiner et al. 2008, PASP 120,
405). Their observations were performed $\sim 40$ min after the GRB,
and the deduced spectral index is $\beta \sim 0.5$ (modeling the
  spectrum as $F_\nu \propto \nu^{-\beta}$), which we assume not to
vary up to the time of our observations. The flux behavior before the
X-shooter observations was estimated using our photometric data from
the IAC80 telescope and other data from the literature (Klotz et
al. 2012a, b, c; Smith \& Virgili 2012; LaCluyze et al. 2012; Malesani
et al. 2012). In detail, if the flux density in the $R$ band is
$F_R=F_R(t_*) \times (t/t_*)^{-\alpha}$, we have $F_R(t_*)=2.1\times
10^{-26}$ erg cm$^{-2}$ s$^{-1}$ Hz$^{-1}$ at $t_*=360$~s, with a
decay index $\alpha = 1.03$.

We chose to model the {\ion{Fe}{ii}} first, because it has a large
number of excited and ground features and thus the best determination
of the column densities. On the other hand, {\ion{O}{i}} and
{\ion{C}{ii}} are useless, since almost all their transitions are
saturated, while {\ion{Ni}{ii}} and {\ion{Si}{ii}} feature just the
ground state and one excited level. We assume that before the GRB
takes place the {\ion{Fe}{ii}} ion is in its ground state, i.e., the
column densities of all the excited levels are zero. At the beginning,
we try to use as the initial column density of the {\ion{Fe}{ii}}
ground state the sum of all the column densities of the {\ion{Fe}{ii}}
levels measured in our spectrum (ground + excited). The exact values
are $\log (N_\ion{Fe}{ii}/{\rm cm}^{-2}) = 15.45 \pm 0.03$ and $\log
(N_\ion{Fe}{ii}/{\rm cm}^{-2}) = 15.49 \pm 0.02$ for components I and
II, respectively. However, these values turn out to be incompatible
with the ratio between the observed excited states. In particular, the
predicted \ion{Fe}{ii} $a^4F_{9/2}$ column density is too high with
respect to the other excited levels, for any reasonable choice of the
GRB/absorber distance and Doppler parameter. For this reason, we
decide to leave the initial column density as a free parameter and
determine through our model the ground state column density
corresponding to the excited level population. Such an approach can
not be applied to the {\ion{Ni}{ii}} and {\ion{Si}{ii}}, since they
feature just two levels, and the model has at least two free
parameters, i.e., the initial column density and the GRB/absorber
distance.

\begin{figure}
\centering
\includegraphics[angle=-0,width=9cm,height=9cm]{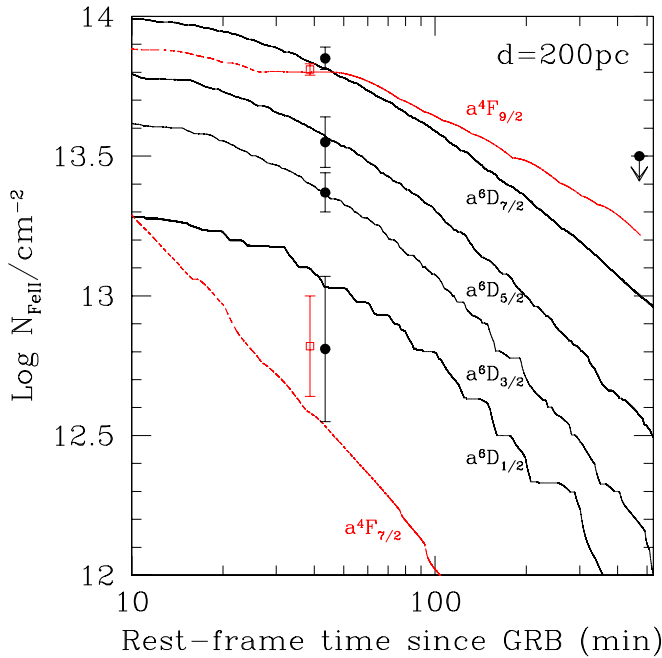}
\caption{{\ion{Fe}{ii}} column densities for the fine structure levels
  of the ground state (black filled circles) and of the first excited
  level (red open squares) for component I in the spectrum of
  GRB\,120327A.  Column density predictions from our time-dependent
  photo-excitation code are also shown.  They refer to the fine
  structure levels of the ground state (solid black lines) and to the
  fine structure levels of the first excited level (red dashed lines),
  in the case of an absorber placed at $200$ pc from the GRB. The
  arrow represents the $90\%$ upper limit to the column density of the
  first fine structure level of the ground state, computed using the
  second-epoch spectrum. For clarity reasons, black points and
    curves have been slightly offset to the right.  }
\label{spe1}
\end{figure}

\begin{figure}
\centering
\includegraphics[angle=-0,width=4.45cm,height=4.45cm]{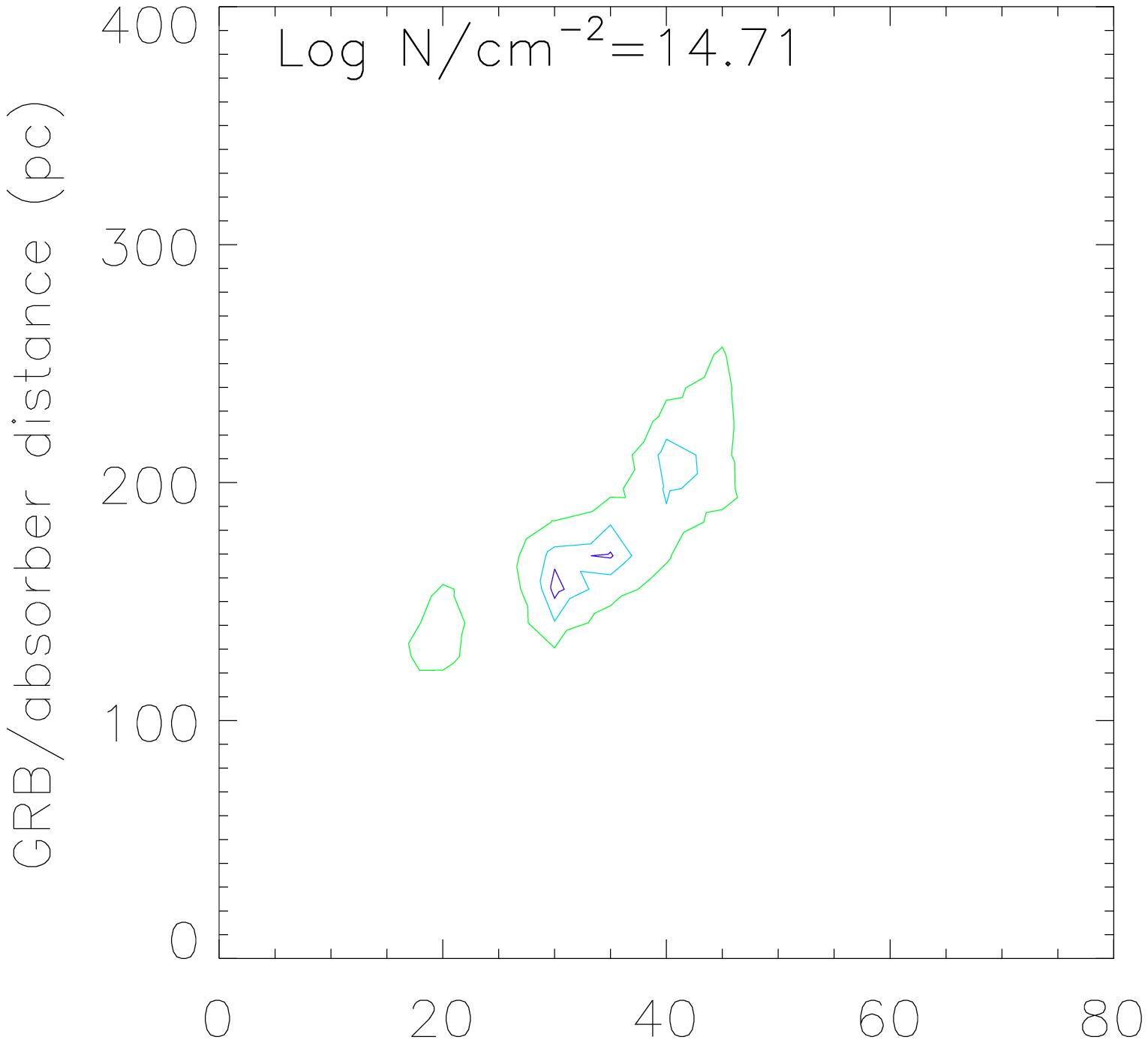}
\includegraphics[angle=-0,width=4.45cm,height=4.45cm]{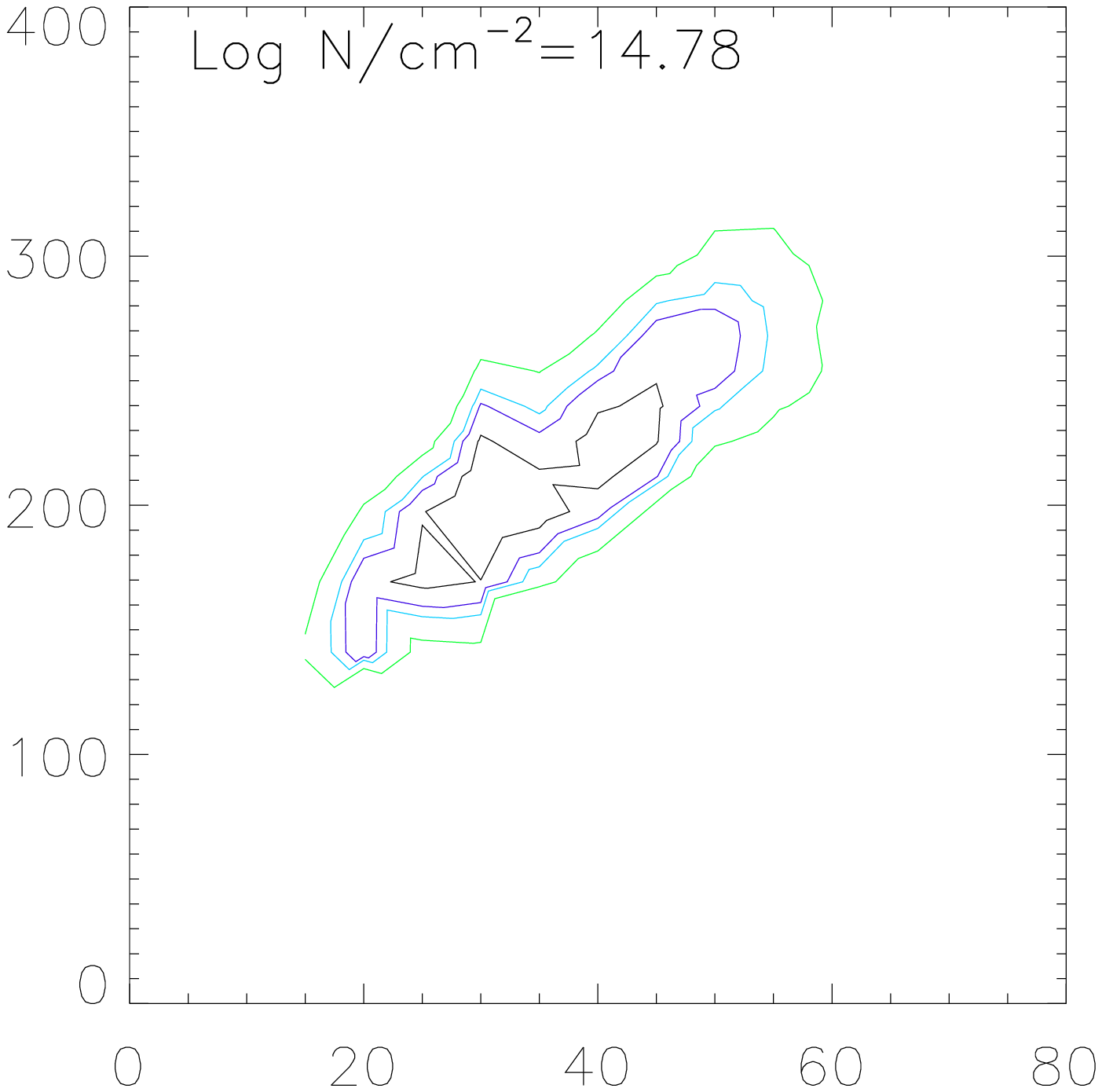}
\includegraphics[angle=-0,width=4.45cm,height=4.45cm]{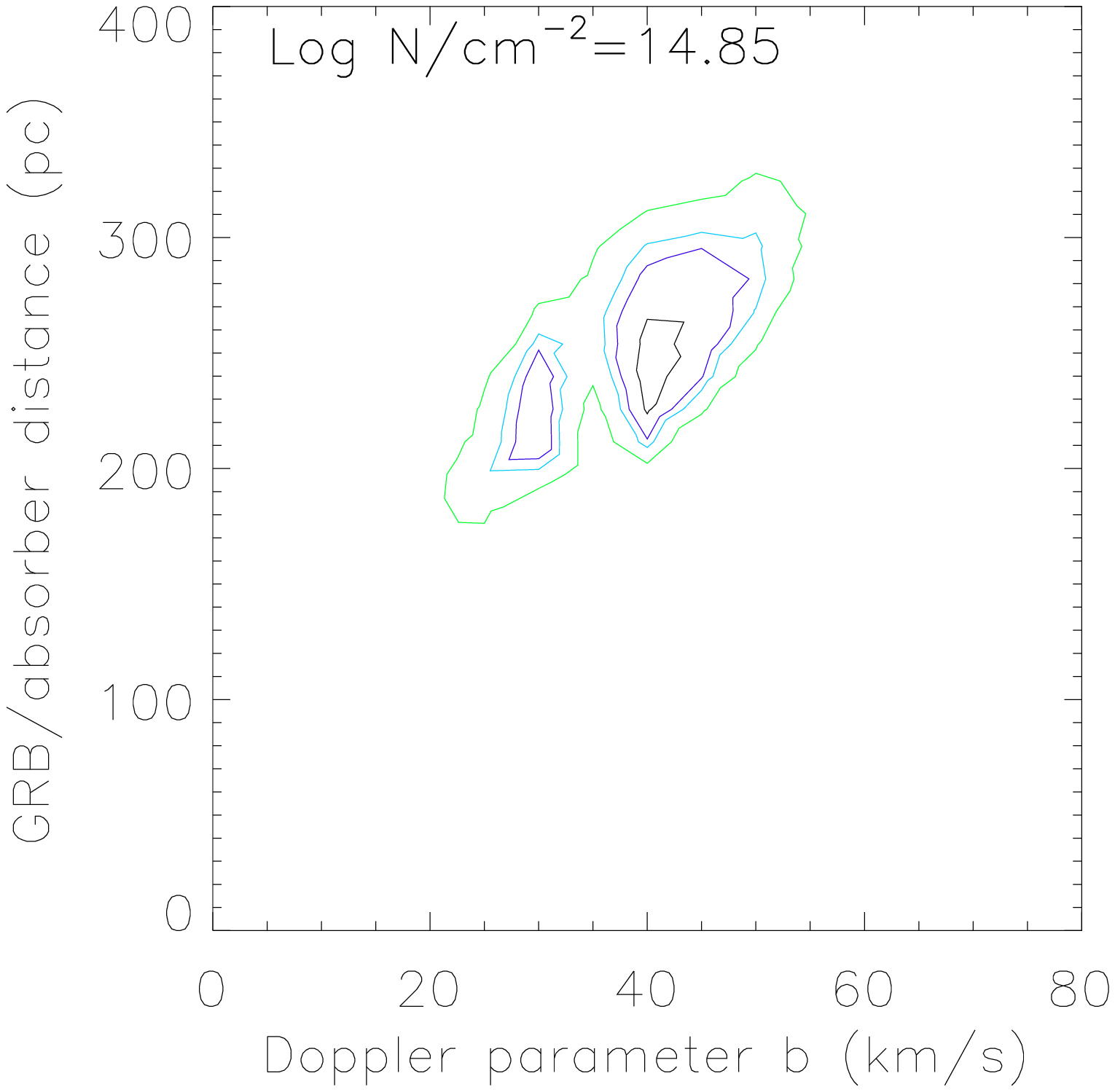}
\includegraphics[angle=-0,width=4.45cm,height=4.45cm]{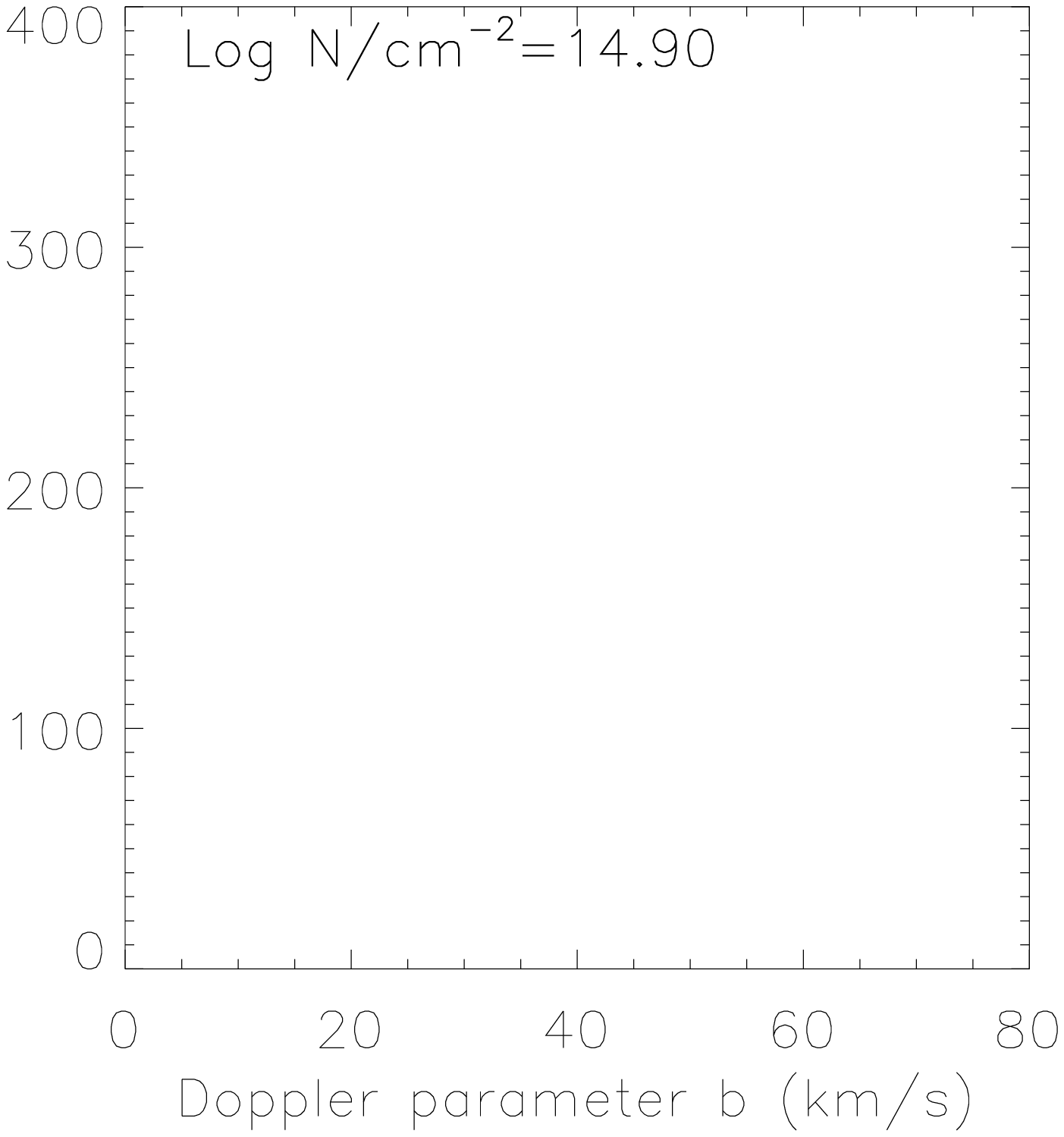}
\caption{The contour plots describing the data/model agreement in the
  computation of the distance between the GRB and component I. The
  four plots display the contours in the plane of Doppler parameter
  and GRB/absorber distance for different initial \ion{Fe}{ii} column
  densities. Black, violet, cyan and green contours enclose the 68\%,
  90\%, 95\% and 99\% confidence regions, respectively. The bottom
  right panel is empty as at the corresponding density no
  combination of Doppler parameter and distance falls inside the 99\%
  confidence surface.}
\label{spe1}
\end{figure}

Fig. 8 shows the best fit model describing the column densities
estimated for component I. The best fit parameters deduced from the
model at the $2\sigma$ confidence level are the following: i) initial
column density (all \ion{Fe}{ii} supposed to be in the ground state
level before the GRB) $\log (N_{\rm I,\ion{Fe}{}}/{\rm cm}^{-2}) = 14.78
\pm 0.07$; ii) Doppler parameter $b_{\rm I}=30^{+25}_{-10}$ km s$^{-1}$;
iii) GRB/absorber distance $d_{\rm I}=200^{+100}_{-60}$ pc. Fig. 9 describes
in more details the data/model agreement showing the contour plots for
component I.

\begin{figure}
\centering
\includegraphics[angle=-0,width=9cm,height=9cm]{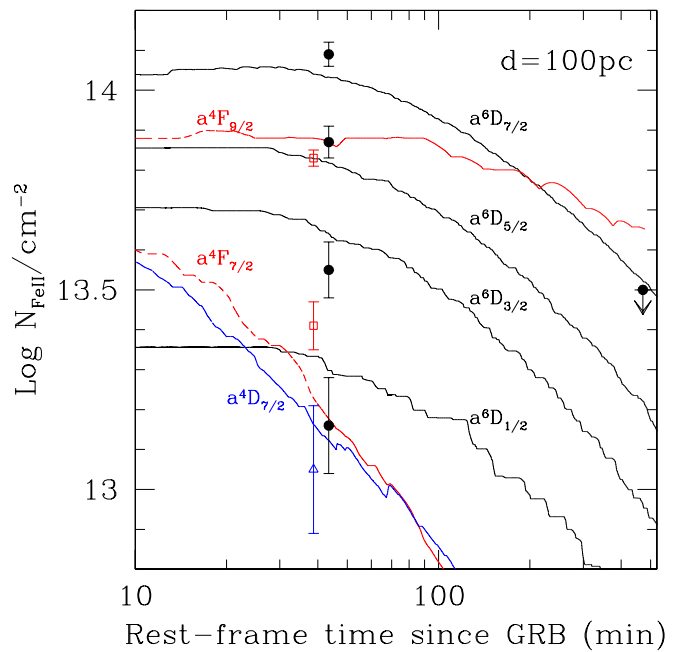}
\caption{Same as in Fig. 8, but for component II and an absorber
  placed at $100$ pc. In addition, the blue, open triangle plots the
  second excited level of the \ion{Fe}{ii}, while the blue solid line
  indicates the corresponding expectation from the time-dependent
  photo-excitation code. For clarity reasons, black points and
    curves have been slightly offset to the right.
}
\label{spe1}
\end{figure}

\begin{figure}
\centering
\includegraphics[angle=-0,width=4.45cm,height=4.45cm]{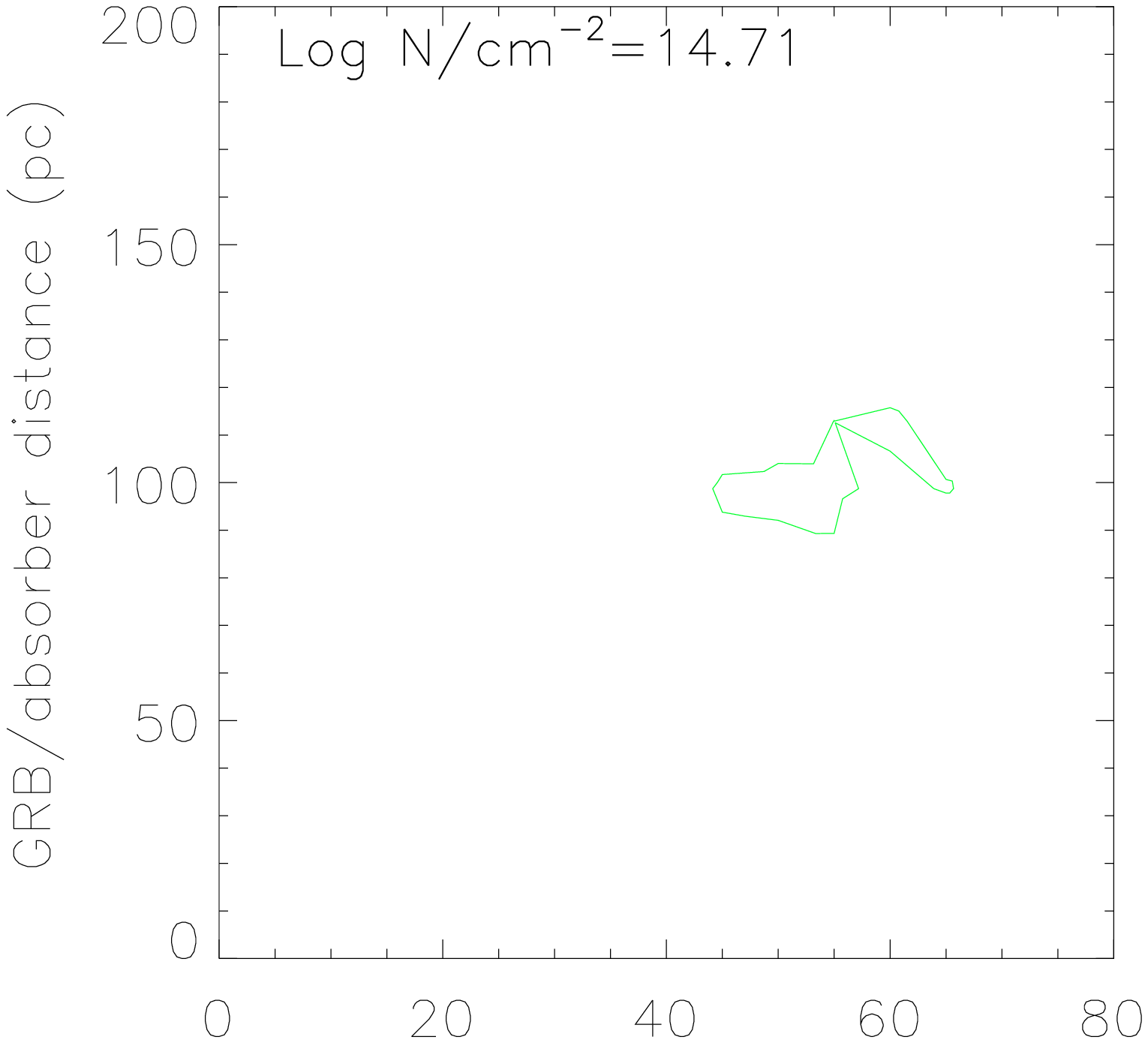}
\includegraphics[angle=-0,width=4.45cm,height=4.45cm]{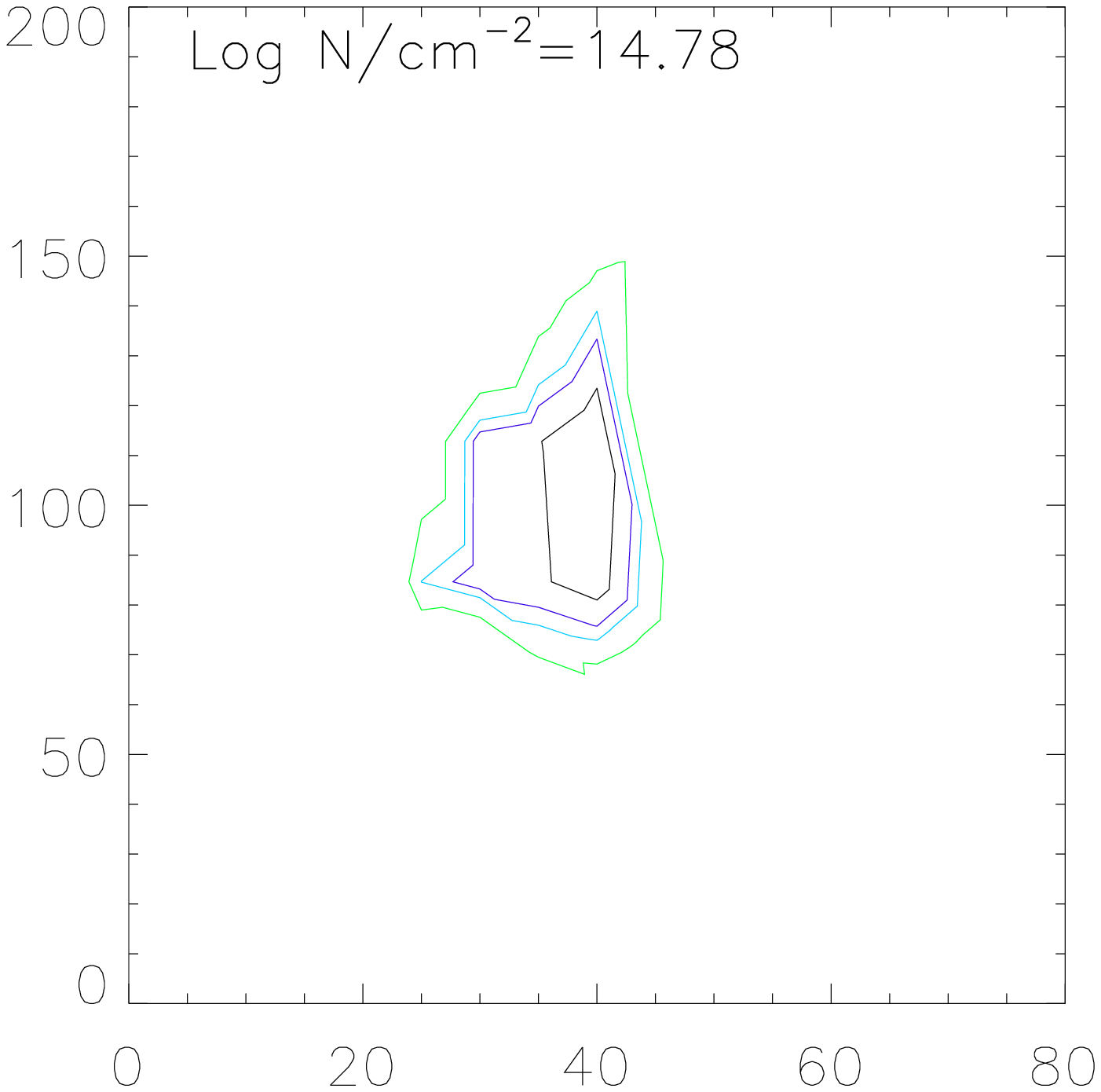}
\includegraphics[angle=-0,width=4.45cm,height=4.45cm]{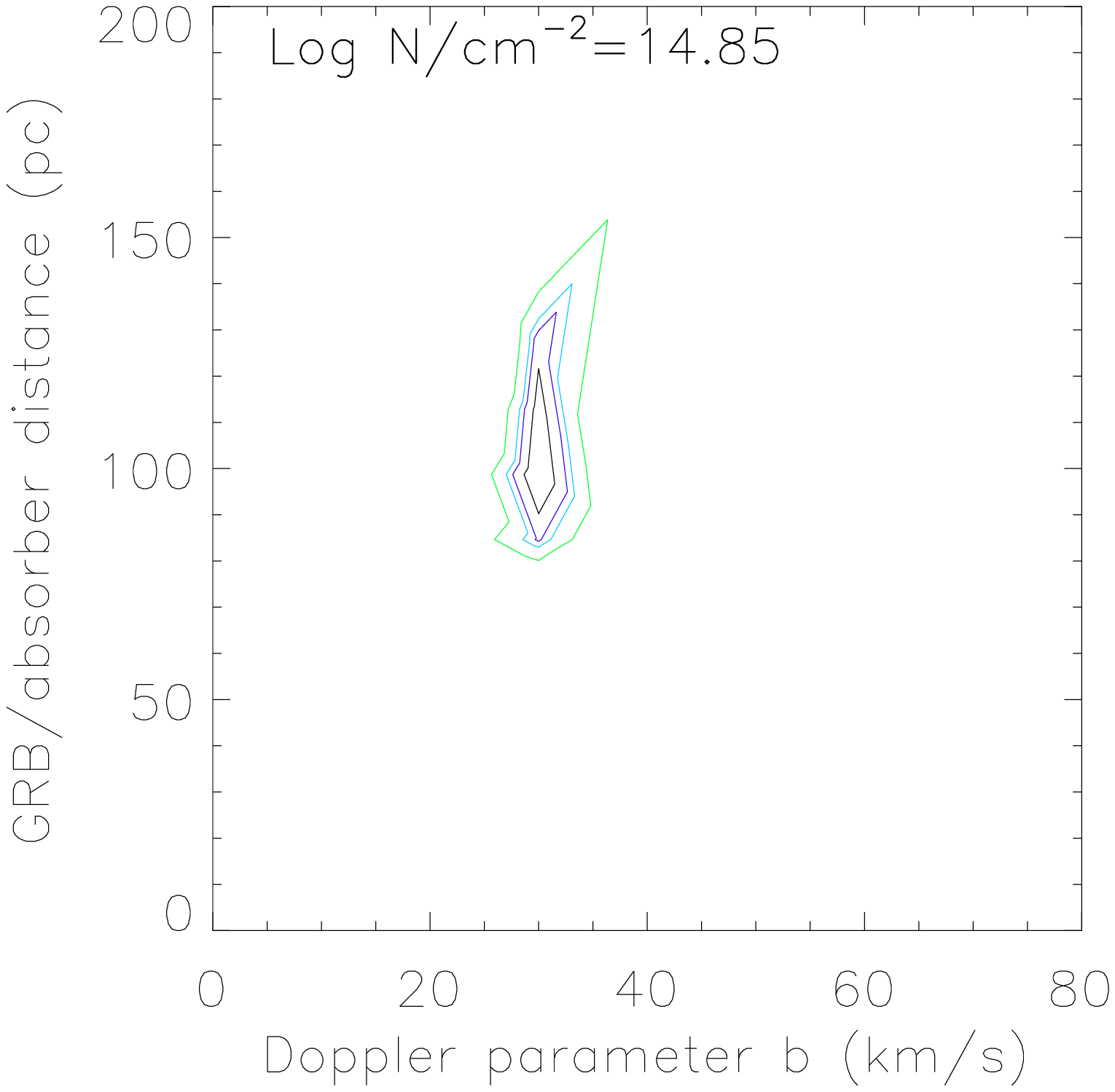}
\includegraphics[angle=-0,width=4.45cm,height=4.45cm]{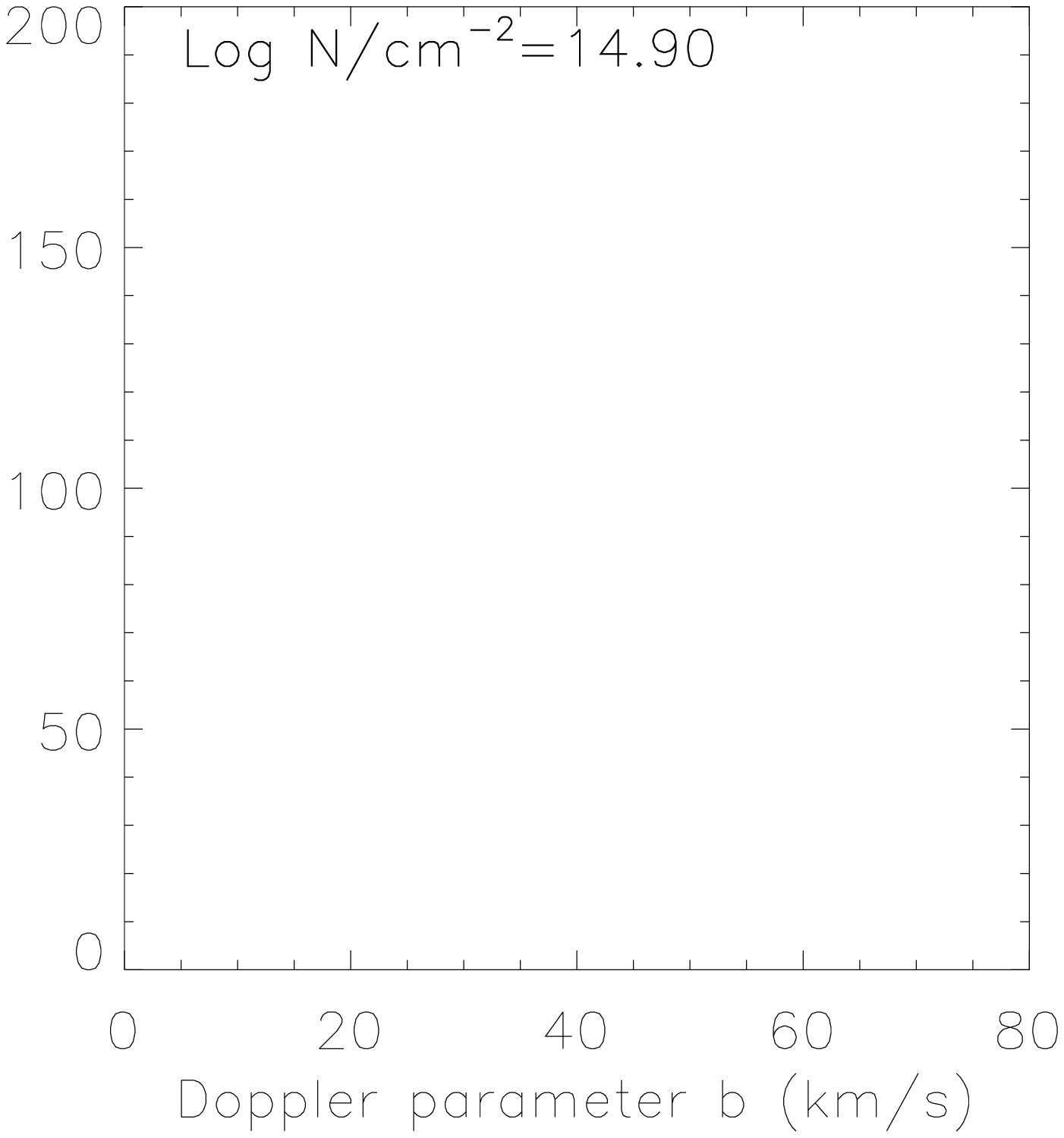}
\caption{The contour plots describing the data/model agreement in the
  computation of the distance between the GRB and component II.  The
  four plots display the contours in the plane of Doppler parameter
  and GRB/absorber distance for different initial \ion{Fe}{ii} column
  densities. Black, violet, cyan and green contours enclose the 68\%,
  90\%, 95\% and 99\% confidence regions, respectively. The
    bottom right panel is empty as at the corresponding density no
    combination of Doppler parameter and distance falls inside the
    99\% confidence surface}
\label{spe1}
\end{figure}

A similar analysis has been performed for component II.  We first try
to model the whole dataset, but we find that the \ion{Fe}{ii}
$a^4F_{7/2}$ level drives the fit towards a high value of the reduced
$\chi^2$, and a GRB/absorber distance which requires high column
densities for the fine structure levels at later times, not in
agreement with the upper limit computed from the second-epoch
spectrum. Fig. 10 shows the best fit model describing the column
densities estimated for component II, not including the \ion{Fe}{ii}
$a^4F_{7/2}$ level. The best fit parameters deduced from the model at
the $2\sigma$ confidence level are the following: i) initial column
density (all \ion{Fe}{ii} supposed to be in the ground state level
before the GRB) $\log (N_{\rm II,\ion{Fe}{ii}}/{\rm cm}^{-2}) =
14.85^{+0.02}_{-0.07}$; ii) Doppler parameter $b_{\rm
  II}=30^{+15}_{-5}$ km s$^{-1}$; iii) GRB/absorber distance $d_{\rm
  II}=100^{+40}_{-30}$ pc. Fig. 11 describes in more details the
data/model agreement showing the contour plots for component II.  If
we include the \ion{Fe}{ii} $a^4F_{7/2}$ level in the fitting
procedure, the GRB/absorber distance is $d_{\rm II}=70^{+20}_{-10}$
pc.



\subsection{Dust depletion patterns}

The observed relative abundances of \ion{Fe}{} and \ion{Zn}{} are
different ([Fe/H]$= -1.73\pm 0.10$ and [Zn/H]$= -1.17 \pm 0.11$). This
is usually attributed to the different refractory properties of the
two elements, with the former that preferentially being locked in
dust grains, and the latter free in the gas phase. The comparison
between these elements is indeed a widely applied tool to derive
information about the dust content along GRB and QSO lines of sight.

Dust depletion patterns can be compared to those observed in
Galactic environments following a technique described by Savaglio, Fall
\& Fiore (2003).  We consider the four depletion patterns observed in
the Milky Way, namely, those in the warm halo (WH), warm disk + halo
(WHD), warm disk (WD) and cool disk (CD) clouds (Savage \& Sembach
1996).

We find that none of the four patterns is able to reproduce or
data (fig. 12). The best fit is given by the WH cloud pattern, with a
metallicity of $\log Z_{\rm GRB}/Z_{\odot} \sim -1.3$ and a GRB
dust-to-metal ratio slightly lower than that of the WH environment in
the Milky Way, e.g., $d/d_{\rm WH} \sim 0.7$. This metallicity value
is consistent with our [Zn/H] measurement as expected, since WH
depletion pattern assumes no depletion of zinc onto dust.

We note that \ion{Si}{} and \ion{Mg}{} are over-abundant compared to
any depletion pattern known in the Galaxy. This can be attributed to
$\alpha$-element over-abundancy with important implications about the
host galaxy nature as discussed in Sect.\,5.3.2. (we note however that
another $\alpha$-element, S, does not appear to be overabundant).

Knowing the hydrogen column density, the metal content and the
dust-to-metal ratio, it is possible to derive an independent estimate
of the extinction (see e.g., Savaglio, Fall \& Fiore 2003).  Assuming
an SMC (Pei 1992) extinction curve recipe we can compute $A_V \sim
0.22$ mag along the line of sight of GRB\,120327A. This is
considerably higher than the value derived in Sect.\,5.5 by modeling
the flux-calibrated spectra ($A_V<0.03$ mag). On the other hand,
hydrogen column densities close to $10^{22}$\,cm$^{-2}$ typically
involve sizable extinctions (e.g. Covino et al. 2013), although the
derivation of the extinction through dust depletion patterns for
GRB\,120327A might not be reliable, given that no Galactic pattern is
able to satisfactorily reproduce our data.


\begin{figure}
\centering
\includegraphics[angle=-90,width=9cm]{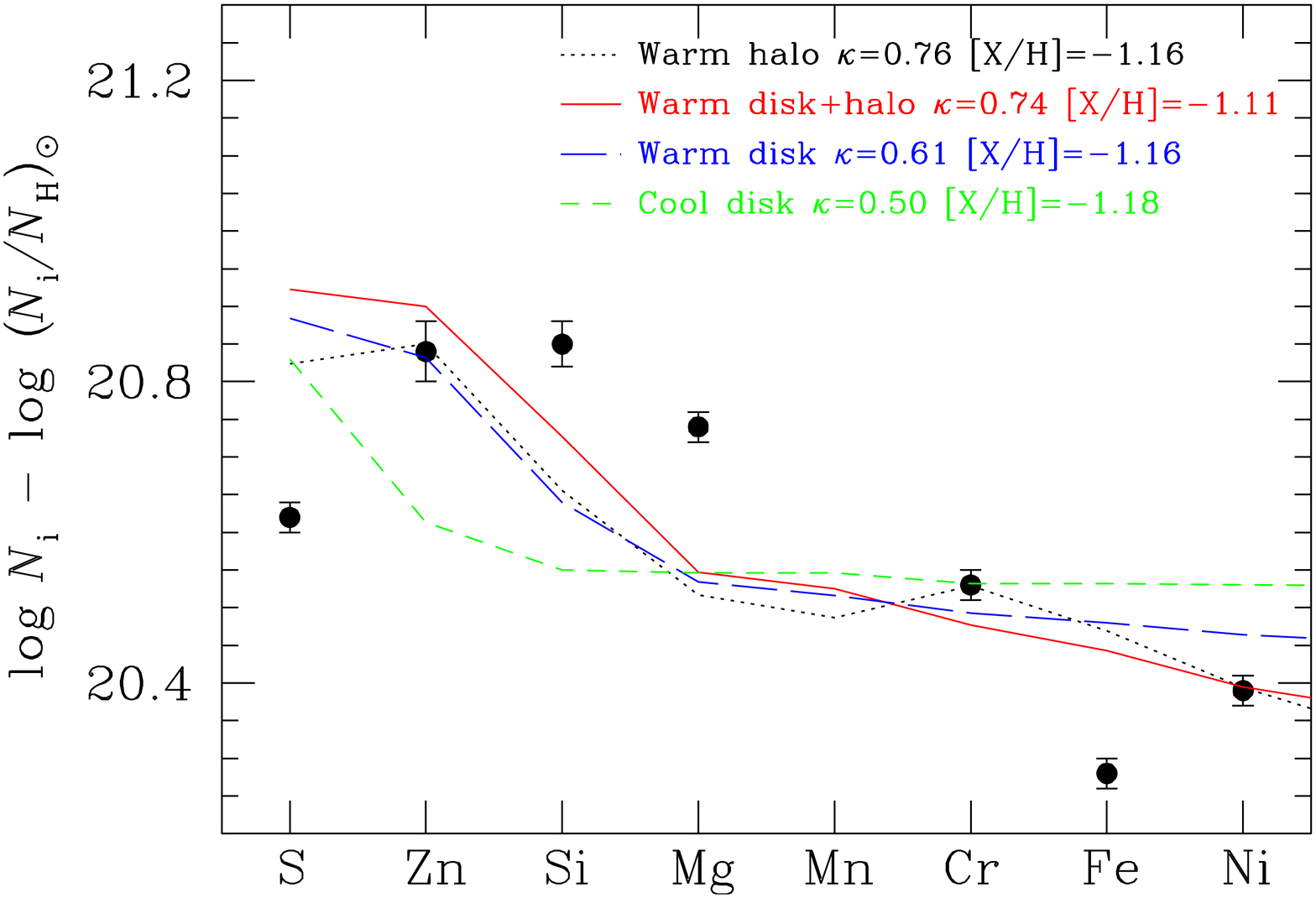}
\caption{Depletion patterns in the absorbing gas of GRB\,120327A,
  taken from average gas-phase abundance measurements in warm halo
  (black dotted line), warm disk + halo (red solid line), warm disk
  (blue long-dashed line) and cool disk (green short-dashed line)
  clouds of the Milky Way (Savage \& Sembach 1996). Filled black
  circles represent our data points, which are best fitted by the warm
  halo cloud pattern. \ion{Si} and \ion{Mg} are clearly
  over-abundandant.}
\label{spe1}
\end{figure}


\subsection{Molecular absorption features}

Weak absorption lines from the Lyman and Werner bands of molecular
hydrogen (H$_2$) are detected at $z=2.8139$ in the X-shooter spectrum
of the afterglow of GRB\,120327A. The observed velocity width of the
lines is similar to the instrumental resolution.  We therefore
constrain the H$_2$ column densities by assuming a range of Doppler
broadening parameters, between 1 and 10 km s$^{-1}$, which should
provide a realistic range on the total H$_2$ column density, i.e.,
$\log (N_{\rm H_2}/$cm$^{-2}) = 15.3$ to $17.7$. This corresponds to
an overall molecular fraction $f = 2\log N_{\rm H_2}/(\log
N_{\ion{H}{i}} + 2\log N_{\rm H-2}) = 4\times 10^{-7}$ to $10^{-4}$.

Details on the column densities in each H$_2$ rotational level are
given in Table 5. $J = 1$, 2 and 3 lines are clearly detected while
for $J=0$ only a few unblended lines are available and they are weak
leading to more uncertain column density estimates. Fig. 13 displays
the two models with $b=1$ and 10 km s$^{-1}$ in blue and red,
respectively. In general, the present measurements are mostly
indicative and should be considered with caution due to the limited
resolution and S/N, and possible blending with Ly$\alpha$ forest
absorption. Only high spectral resolution observations are suitable to
properly resolve the H$_2$ lines, measure accurately the column
densities, and study the physical conditions in the gas (Srianand et
al 2005; Noterdaeme et al. 2007). Nevertheless, the presence of H$_2$
in this system is firmly established. This constitutes the third clear
detection of H$_2$ absorption in a GRB afterglow spectrum (Prochaska
et al. 2009, Kr\"uhler et al. 2013). It is interesting to note that the
molecular lines are lined up with the bluemost component I.

\begin{table}[ht]
  \caption{\bf H$_2$ column densities towards GRB\,120327A}
\centering
{\footnotesize
\smallskip
\begin{tabular}{|l|c|c|}
  \hline 
  Rot. level & \multicolumn{2}{c|}{$\log N_{\rm H_2}/{\rm cm}^{-2}$} \\
  component  & $b=1$ km~s$^{-1}$ & $b=10$ km~s$^{-1}$   \\
  \hline                          
  \hline                          
  $J=1$         & $17.6$                 & $15.1$            \\
  $J=2$         & $16.0$                 & $14.2$            \\
  $J=3$         & $17.1$                 & $14.8$            \\
  \hline  
  Total         & $17.7$                 & $15.3$            \\
  \hline

\end{tabular}
}
\end{table}

\begin{figure}
\centering
\includegraphics[angle=0,width=9cm]{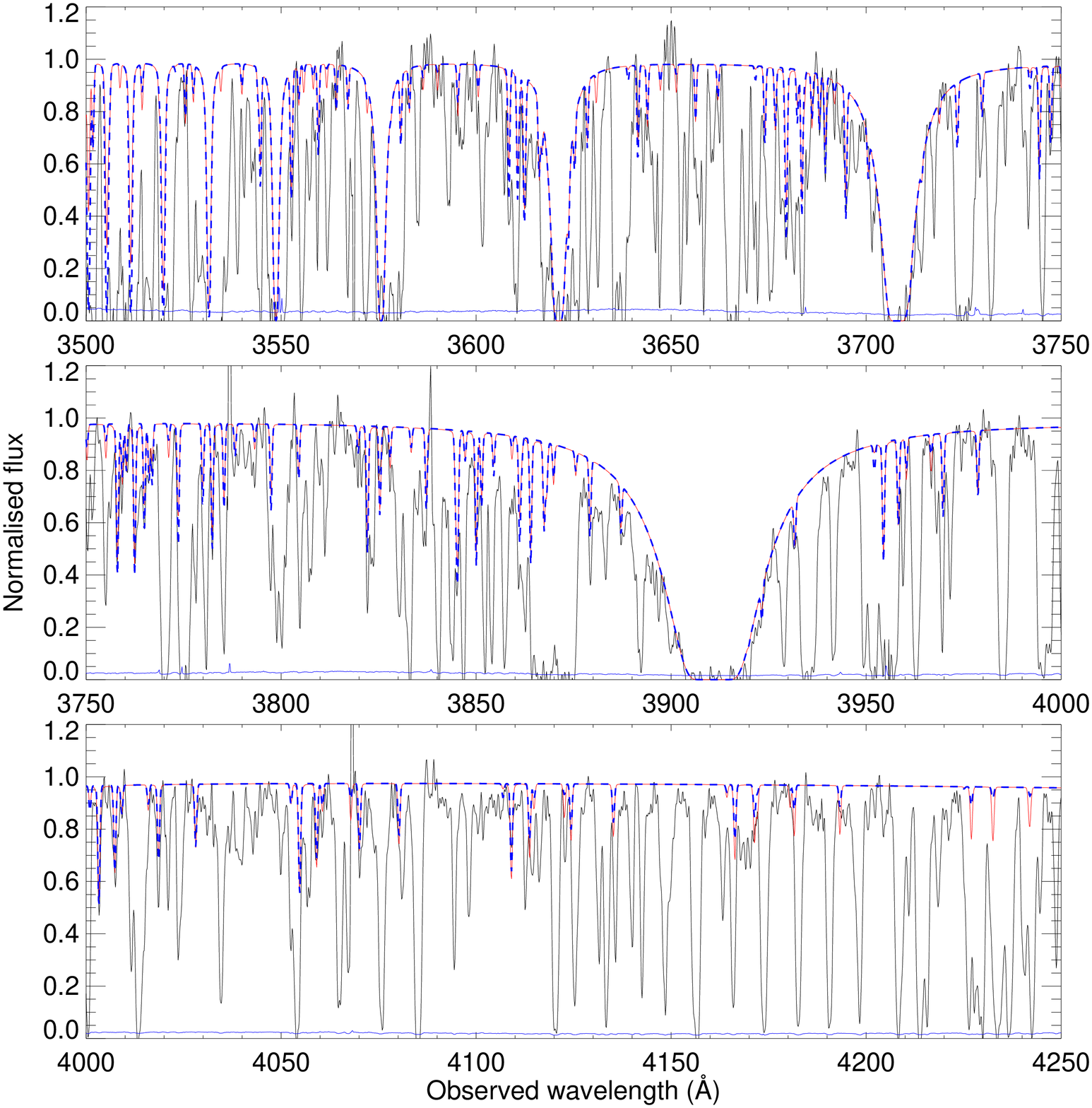}
\caption{The single component fit for the H$_2$ absorption lines. The
  two models with $b=1$ and $b=10$ km s$^{-1}$ are shown in dashed
  blue and solid red, respectively. These models are barely
  distinguishable by eye, a deep zoom is necessary to catch the
  different profiles. The X-shooter spectrum is overimposed in
  black. The bottom, blue solid line displays the error spectrum.}
\label{spe1}
\end{figure}

\subsection{The extinction curve shape}

The shape of extinction curves at high redshift is a powerful test for
deriving information about dust formation and possibly about the
various processes affecting dust absorption and destruction close to
GRB sites. In most cases, inferences about dust extinction curves are
obtained by modeling the spectral energy distribution obtained through
photometric observations of GRB afterglows, which unfortunately
suffers from a strong degeneracy between afterglow spectral slope and
extinction. Moreover, for very high redshift events in particular,
uncertainties in photometry of single bands, which can also present
specific calibration problems may affect the whole analysis (see for
instance the discussion in Stratta et al. 2007 and Zafar et al. 2010
about GRB\,050904). For early time afterglows, the lack of
simultaneity (within few hours) of the available photometric
information has to be properly considered to avoid spurious results.

For most GRB afterglows, when accurate multi-band photometry is
available (e.g., Covino et al. 2008; Schady et al. 2010; Kann et
al. 2010; Covino et al. 2013), the derived extinction curve is in fair
agreement with what is observed locally in the Small Magellanic Cloud
(SMC, Pei 1992), although often, because of the limited wavelength
resolution, this simply means that the observed extinction curve has
to be chromatic (i.e. wavelength dependent) and featureless. A few
remarkable exceptions have been recorded in an intervening system
along the line of sight of GRB\,060418 (Vreeswijk et al. 2007), for
GRB\,070802 (Kr\"uhler et al. 2008; El\'{\i}asd\'ottir et al. 2009)
and GRB\,080607 (Prochaska et al. 2009), where the characteristic
absorption feature at about 2175\,\AA, prominent in the Milky Way
extinction curve, has also been detected (see also Zafar et al. 2012
for a review on this feature in GRB hosts).

Clearly, when spectroscopic information is available (e.g., Watson et
al. 2006; El\'\i asd\'ottir et al. 2009; Liang \& Li 2010, Zafar et
al. 2011), the extinction effect and intrinsic spectral slope can be
better disentangled. The case of GRB\,090926A is a good example of the
X-shooter capabilities in this context (D'Elia et al. 2010). Moreover,
a paper exploiting the full X-shooter GRB sample to address these
issues is currently in preparation (Watson et al. 2014). In the case of
GRB\,120327A, however, because of the malfunctioning in the
X-shooter ADC, our flux calibration is reliable only in the NIR arm,
where atmospheric dispersion is much less important and no such device
is necessary. In order to extend our spectroscopic coverage to optical
wavelengths, we used the nearly simultaneous, flux-calibrated GTC
spectrum.

We removed wavelength intervals affected by telluric lines of strong
absorptions. Since the Ly$\alpha$ range is important for extinction
determination, we have included it in the analysis using the hydrogen
column density reported in Table\,2. We rebinned the spectrum in bins
of approximately 50\,\AA$\;$ by a sigma-clipping algorithm to avoid
the effect of residual absorption systems.  Assuming a power-law
spectrum, we obtained $\beta \sim 0.5$ (consistent with the value
  deduced by the GROND data reported in Sudilovsky et al. (2013)),
where the afterglow spectrum is modeled, as is customary in GRB
literature, as $F_\nu \propto \nu^{-\beta}$. We tried to fit the data
using different extinction curves, namely, SMC, LMC and Milky Way. The
best fit is obtained with an extinction of $E_{B-V} < 0.01$ mag
($A_V<0.03$ mag) at $95\%$ confidence level, regardless of the adopted
curve (although a higher extinction could be inferred by the metal
column densities, as discussed in Watson et al 2013). We caution that
possible systematics in the absolute calibrations of the various
spectra involved in the analysis could affect the results well beyond
the statistical uncertainties here reported.

\section{Discussion}

\subsection{The GRB\,120327A line of sight}

We have shown that the absorption lines in the X-shooter spectrum of
the afterglow of GRB 120327A are well described by a three component
Voigt profile (see Table 2).  Strangely enough, the three detected
features are distributed across a narrow velocity interval ($\approx
80$ km s$^{-1}$). Given that the host galaxy is particularly bright
and rich in metal lines, we could expect a large mass, which
should correspond to a large velocity dispersion.

It is interesting to note that different species show different
abundance ratios among the three components. Low ionization species
show absorption in all components, with the redmost component III
being less populated than components I (bluest) and II, which in turn
feature a similar column density. The only exception is represented by
\ion{Ca}{ii} and \ion{P}{ii} which do not present absorption in
component III, with the former featuring also a higher column density
in component I.

On the other hand, neutral species do not show absorption in component
III (with the exception of \ion{Ar}{i} and the saturated \ion{O}{i}),
and component I has generally a column density which is higher than
that of component II. This is particularly evident in \ion{Ca}{i} and
\ion{P}{i}, where we can only set upper limits for the component II
column densities. Again, \ion{Ar}{i} does not follow this behaviour,
featuring a comparable absorption in the two components.

These data can be easily explained assuming that component II is
closer to the GRB explosion site than component I. In fact, in this
scenario, part of the absorbing gas in component I can survive as
neutral atoms because its distance from the GRB radiation is large
enough to avoid complete ionization of the atomic species. The larger
column density of component I of \ion{Ca}{ii} with respect to
component II is explained because this species has an ionization
potential of just $11.9$ eV, thus smaller than that of
\ion{H}{i}. This means that the hydrogen is transparent to a large
portion of photons able to ionize \ion{Ca}{ii} to \ion{Ca}{iii} (those
with energy between $11.9$ and $13.6$ eV), and is not able to screen
this transition.

Similarly, neutral species show a fainter absorption (if any) in the
second component with respect to the first one due to their ionization
potential being lower than that of hydrogen. 
\ion{Ar}{i} is an exception, because of the comparable column
densities for components I and II. This is not unexpected, since
\ion{Ar}{i} has the highest ionization potential among the neutral
species, well above that of \ion{H}{i}. This scenario is confirmed by
the $H_2$ molecule being detected in component I, and not in component
II, where the GRB flux is possibly too high to allow $H_2$ to survive.

This scenario is confirmed by the determination of the GRB/absorber
distances using the excited line variation and ratios: component I is
$\sim200$ pc away from the GRB, component II $\sim100$ pc. These
distances are among the shortest computed using this kind of analysis,
but fully consistent with other determinations (see Hartoog et
al. 2013). Moreover, the marginal or absent detection in component II
of many neutral species with a low ionization potential is not
unexpected. In fact, the GRB flux should fully ionize \ion{Mg}{i} into
\ion{Mg}{ii} on a distance scale of $50-100$ pc (Prochaska et
al. 2006).

The comparison between the \ion{Fe}{ii} excited levels and a
time-dependent photo-excitation code gives us another piece of
information about the gas geometry around GRB\,120327A. In fact, the
modeling requires less ground state \ion{Fe}{ii} column densities both
in component I and II than those observed, in order to reproduce the
behaviour of the excited levels. This is a possible indication that
only a fraction of \ion{Fe}{ii} resides at $100$ pc and $200$ pc from
the GRB. In other words, most of the gas detected in component I and
II could be located far away from the GRB, in a region where the
\ion{Fe}{ii} excited features had enough time to de-excite into the
ground level when the X-shooter observation began. This statement is
supported by the absorption spike present in component I of some
lines, such as {\ion{Si}{ii}$~\lambda$1808},
{\ion{Al}{iii}$~\lambda$1854} and {\ion{Fe}{ii}$~\lambda$2260} (see
fig. 2 and 3), a feature that can not be closely matched by our fits
performed at the nominal X-shooter resolution.


Several high ionization lines have been detected (fig. 6, Tab. 2). The
best fit is obtained with a single Voigt profile, centered exactly in
the middle between components I and II. The only exception is
represented by \ion{N}{v}, whose profile is nicely lined-up with
component II. A more detailed modeling is not possible, since many of
these lines are saturated or fall in the UVB arm, which has a lower
resolution. However, the \ion{N}{v} fit might suggest that high
ionization lines tend to prefer component II, the closer to the GRB,
maybe due to a strong ionization effect (see De Cia et al. 2012;
Vreeswijk et al. 2013), but this is definitely not a conclusive
argument.

Concerning the shape of the high ionization transitions, Fox et
al. (2008) showed that, on average, they present blue wings or high
velocity components with respect to low ionization species, suggesting
that they may arise in a separate phase. This different morphology
does not arise in the absorption features of GRB\,120327A, with the
possible exception of the \ion{O}{vi} lines, which are however heavily
saturated. Prochaska et al. (2008) noted that GRB sightlines show more
frequent and much stronger \ion{N}{v} than random sightlines through
high redshift galaxies and that this, together with a line profile
similar to that of fine structure lines, hints toward a circumburst
origin of the \ion{N}{v}. They concluded that if the latter is
produced by photoionization of \ion{N}{iv} by the burst radiation, it
has to reside at distances of the order of $10$ pc from the GRB and
show strong variability. This turned out not to be the case for
GRB\,080310, for which time resolved high resolution spectroscopy was
available. For this GRB, a foreground host galaxy ISM scenario is
favoured for the production of \ion{N}{v} (De Cia et
al. 2013). However, the \ion{N}{v}, \ion{C}{iv} and \ion{Si}{iv} column
densities are still way lower than what observed in X-ray spectra. The
X-ray absorption is likely dominated by the circum-burst medium, where
the gas is extremely ionized, much more than where \ion{N}{v},
\ion{C}{iv} and \ion{Si}{iv} are (see Watson et al. 2013; Schady et
al. 2011). For what concerns GRB\,120327A, the S/N of the second
X-shooter observation is not sufficient to assess the variability of
\ion{N}{v}.


\subsection{The GRB\,120327A host galaxy}

Relative abundances of the GRB\,120327A host galaxy have been computed
for all the species detected in absorption (see Table 4). While
performing this computation, we did not take into account high
ionization lines. 
Nevertheless, the abundances computed including or not the high ions
do not differ much, because high ionization column densities are much
smaller than lower ionization ones for GRB\,120327A (not considering
the extremely saturated oxygen). The element for which this difference
is higher is Phosphorus. In this case, $\log N/{\rm cm}^{-2} =
14.19\pm0.04$ ($14.25\pm0.04$) excluding (including) the {\ion{P}{v}}
column density.

The metallicity computed from the less refractory elements is between
$-1.4$ and $-1.1$ with respect to solar values (see Table 4). This
value lies in the middle of the metallicity distribution of GRB host
galaxies (Savaglio 2012), which extends from [X/H] less than $-2$ as
in GRB\,090926A at $z=2.11$ (Rau et al. 2012; D'Elia et al. 2012), up
to super solar values as in GRB\,090323 at $z=3.57$ (Savaglio et
al. 2012).

Using our late time GTC photometry, the detection of the host in the
$r$ band can be used to derive the star-formation rate, not corrected
for host dust extinction (Savaglio, Glazebrook \& LeBrogne
2009). Using their equation 5, a magnitude of $r' = 24.5$, translates
into a lower limit for the star-formation rate of $\mbox{SFR} \sim 5\,
M_\odot$ yr$^{-1}$. 

Concerning the GRB\,120327A host characterization, the $\alpha$
element enhancement with respect to iron group elements points toward
a massive galaxy. In addition, if we correct our $r'$ band late-time
photometry for Galactic reddening, we obtain $r'\sim 23.7$ for the
host galaxy. This is quite a bright value for a GRB host at
$z=2.8$. For example, GRB\,021004 at $z=2.33$ has $R=24.1$ and $\log
M/M_{\odot}\sim10.2$ (Savaglio et al. 2009). Indeed, Hjorth et
  al. (2012) showed that none of the TOUGH hosts at $2<z<3$ is brighter
  than $R \sim 24$ (Vega). Finally, a comparison with chemical
evolution models (Matteucci 2001; Calura et al. 2008, 2009; Grieco et
al. 2012) favours a strong star formation efficiency ($~10$ Gyr$^-1$)
in the GRB host, with a value similar to those typical of massive
proto-spheroids (e.g., Pipino et al. 2011).  It is worth to stress
that single GRB-DLA studies sample only a single sightline through the
host galaxy, and that different physical conditions may characterise
other sightlines. However, by extending this study to a larger set of
GRB hosts, in principle it could be possible to have more precise
information on the global chemical abundance pattern and on the star
formation history of such systems as a population. This task is
deferred to future work and is already feasible, as more and more new
abundances from GRB absorption studies have become available in the
last few years. Even more will become available thanks to next
generation of medium/high resolution spectrographs, such as ESPRESSO
and ELT-HIRES, instruments optimised to perform absorption line
studies at best and which, in principle, will offer the opportunity to
probe various lines-of-sight of even distant, single objects. In this
framework, the utility of chemical evolution models as unique tools to
gain fundamental information on a few basic, fundamental parameters of
GRB hosts such as their nature and ages, should be seriously
considered.

Apparently, the aforementioned massive galaxy as the host for
GRB\,120327A is against the claim that most of the GRB hosts are
small, star-forming, blue and metal-poor galaxies (see, e.g.,
Vreeswijk et al, 2001; Christensen et al. 2004; Fruchter et al. 2006;
Wiersema et al. 2008, Levesque et al. 2010). However, this picture is
based on samples constituted mainly by low redshift galaxies
($z<2$). Recent, deep observations seem to suggest that this could be
only a partial view, and that at high redshift the situation might be
different (Kr\"uhler et al. 2011; Hunt et al. 2011; Savaglio et
al. 2012; Elliot et al. 2013, Perley et al. 2013). As suggested by
Savaglio (2012), the key role of this picture must be played by the
star formation, which appears to be strong in most of the GRB host
galaxies. GRBs are associated with the death of massive and
short-lived stars ($M>30 M_{\odot}$, see e.g. Heger et al. 2003),
which in turn must reside in highly star-forming environments. In
addition, star-formation in the local universe mainly occurs in small,
blue galaxies, i.e., the typical GRB hosts at low $z$. On the other
hand, most of the star-formation at higher redshift is likely to occur
in more massive galaxies (Juneau et al. 2005; Hunt et al. 2011). Thus,
it is not implausible to expect that some GRB hosts at $z>2$ could be
high mass galaxies, given also that many dark GRB hosts are massive
(e.g. Berger et al. 2007; Levan et al. 2006; Svensson et al. 2012;
Perley et al. 2013). Our result, which seems to point towards a high
mass for the GRB\,120327A host galaxy, confirms that the morphology of
high redshift GRB hosts may not resemble that of low redshift ones.

\subsection{Dust content and molecules}

The formation of stars is tightly correlated with the presence of
H$_2$ (e.g., Shu et al. 1987; Blitz 1993; McKee \& Ostriker 2007;
Bigiel et al. 2008). Thus, molecular gas is expected to be found in
star-forming environments such as GRB surroundings, but the search for
its absorption features has often given negative results (see
e.g. Tumlinson et al. 2007). Fynbo et al. (2006) interpreted an
absorption feature of the GRB\,060206 afterglow spectrum as a possible
H$_2$ detection. Prochaska et al. (2009) and Sheffer et al. (2009)
report the presence of strong H$_2$ and CO absorption features in the
spectrum of the GRB\,080607 afterglow. More recently, Kr\"uhler et
al. (2013) reported a strong absorption of H$_2$ in the X-shooter
spectrum of GRB\,120815A, at $z=2.36$, finding a molecular fraction of
$f=0.07$.

For GRB\,120327A we found $f=5\times 10^{-7} - 10^{-4}$, which is the
third detection of molecules in a GRB environment. The molecular
features are perfectly lined up with component I. This, however, is
not enough to assess that the molecules are located at a distance of
$\sim 200$ pc from the GRB, because, as noted throughout the paper,
both components I and II receive contribution in absorption also from
gas which is far away from the GRB. The molecular fraction in
GRB\,120815A is more than two orders of magnitude higher than what we
found for GRB\,120327A, although the two GRBs feature an almost
identical, large $N_{\ion{H}{i}}$ column density of $\log (N_{\rm
  H}/{\rm cm}^{-2}) \sim 22$ and metallicity ($\mbox{[Zn/H]} \sim
-1.1$). More surprisingly (H$_2$ is generally stronger in dust rich
environments), our burst appears not to have a high dust content.
For GRB\,120327A we measure $\mbox{[Zn/Fe]} = 0.56\pm0.14$, whereas, for
GRB\,120815A, Kr\"uhler et al. (2013) report a value of
$\mbox{[Zn/Fe]}=1.01\pm0.10$, which means that a higher quantity of dust is
present along the line of sight of this GRB. This is important,
because dust plays a major role for the presence of molecular gas by
catalyzing the formation of molecules on the surface of dust grains as
well as shielding against Lyman-Werner photons. Guimaraes et
al. (2012) studied the QSO-selected system SDSS J081634+144612 at
$z=3.287$, which shows nearly identical $N_{\ion{H}{i}}$, [Zn/H] and [Zn/Fe]
($\log (N_{\rm H}/{\rm cm}^{-2}) = 22.0\pm 0.1$,
$\mbox{[Zn/H]} = -1.10\pm0.10$,
$\mbox{[Zn/Fe]} = 0.48\pm0.06$). They detect an H$_2$
molecular fraction of $f \sim 9 \times 10^{-4}$, a value closer to
that of GRB\,120327A. Noterdaeme et al. (2008) reported that in
QSO-DLAs the Iron locked into dust ($N_{\rm Fe,dust}$) is a primary driver
of H$_2$ detection, and that $N_{\rm Fe,dust}$ can be linked to a
combination of metallicity, dust depletion and column density. Our
measurement thus confirms the importance of dust in the molecular
formation.

 \section{Conclusions}

In this paper we present multi-epoch, intermediate-resolution
($R=7000 - 12000$) spectroscopy of the optical afterglow of
GRB\,120327A, observed using the X-shooter spectrograph at the VLT,
$\sim 2.5$ and $27.7$ hr after the trigger. 

Our main findings are summarized below.

(i) From the detection of hydrogen and metal absorption features, we
find that the redshift of the host galaxy is $z=2.8145\pm0.0001$. The
ISM of the GRB host galaxy has at least three components contributing
to the absorption. The line center of component I (III) is $41$ ($35$)
km s$^{-1}$ bluewards (redwards) of component II, which sets the
redshift reference point.

(ii) Component III presents less absorption features and no excited
lines, and is probably the farthest from the GRB explosion site.

(iii) Components I and II have similar absorption profiles, but also
some differences, with the former showing more absorption from neutral
elements and the latter from highly ionized species. Fine structure
and metastable levels are present in both components, but with higher
column densities in component II. This points towards component II
being closer to the GRB than component I.

(iv) High ionization lines do not present blue wings or higher
velocity components with respect to low ionization ones, but show a
similar velocity range, contrary to what is observed in previous GRBs.

(v) The distance of these components from the GRB is computed using a
time dependent photo-excitation code and comparing the results with
the excited line ratios for the \ion{Fe}{ii} ion. Component I is
$200^{+100}_{-60}$ pc away from the GRB explosion site, while
component II is at $100^{+40}_{-30}$ pc, confirming result (iii). The
variability of the \ion{Fe}{ii}~$\lambda$2396 fine structure line
between the two X-shooter spectra proves that the excited levels are
produced by the GRB UV flux.

(vi) The column density of the \ion{Fe}{ii} ground state is in excess
from what is expected using the photo-excitation code, both in component I
and II. This possibly indicates that most of \ion{Fe}{ii} lies far
away from the GRB, and that the geometry of the GRB\,120327A host
galaxy absorber is more complex with respect to the resolving capacity
of X-shooter.

(vii) The damped Lyman-$\alpha$ system associated with the GRB host
galaxy is prominent, with a column density of $\log (N_{\rm H}/{\rm
  cm}^{-2}) = 22.01 \pm 0.09$. The metallicity computed using the less
depleted elements is between $\mbox{[X/H]} = -1.3$ and $-1.1$, i.e., a
value in the middle of the metallicity distribution for the GRB host
galaxies at high redshift.

(viii) The dust depletion pattern indicates an $\alpha$ element enhancement
with respect to the elements of the Iron group, typical of massive
galaxies. The same conclusion is supported by modeling this
overabundance using detailed chemical enrichment codes.

(ix) A joint analysis of host galaxy photometry and chemical
enrichment codes leads us to suggest that the host is a high mass
($>10^{10}$ $M_{\odot}$), high luminosity one.

(x) Concerning the dust content in the host, we estimate $A_V = 0.22$
mag along the line of sight of GRB 120327A, using the dust depletion
pattern, considerably in excess of what we find studying the
extinction curve shape by modeling the flux-calibrated spectra ($A_V <
0.03$ mag). The ratio $\mbox{[Zn/Fe]} = 0.56\pm0.14$ depicts an
environment with a certain dust depletion, but not so extreme as in
other dusty GRBs such as GRB\,120815A or GRB\,080607.

(xi) The H$_2$ molecule is firmly detected in the GRB\,120327A host
galaxy. This is the third secure detection of molecules in a GRB
environment after GRB\,080607 and GRB\,120815A. The molecular fraction
we determined is at least two orders of magnitude below that of the
other GRBs, possibly because of the lower dust content, since it helps the
molecule formation.

\begin{acknowledgements}
  We thank an anonymous referee for several helpful comments that
  improved the quality and clarity of the paper. We thank Eros
  Vanzella for several helpful discussion. VDE acknowledges partial
  support from PRIN MIUR 2009. JPUF and DX acknowledge support from
  the ERC-StG grant EGGS-278202. AdUP acknowledges support by the
  European Commission under the Marie Curie Career Integration Grant
  programme (FP7-PEOPLE-2012-CIG 322307). The research activity of
  AdUP and JG is supported by Spanish research project
  AYA2012-39362-C02-02. EP acknowledges INAF PRIN 2011 and ASI INAF
  contract I/088/06/0. The Dark Cosmology Centre is funded by the
  Danish National Research Foundation. The 0.82m IAC80 Telescope is
  operated on the island of Tenerife by the Instituto de Astrof\'isica
  de Canarias in the Spanish Observatorio del Teide.  We acknowledge
  Christophe Martayan (ESO) for assistence to address the
  malfunctioning of the X-shooter ADCs.
\end{acknowledgements}

\appendix
\section{IAC-80 photometry}

In this appendix we present the IAC-80 data, used to produce the lower
panel of Fig. 1.

\begin{table}[ht]
  \caption{IAC-80 photometric I-band observations in the Vega system}
\centering
\smallskip
\begin{tabular}{|l|c|c|c|}
  \hline
  Hours from GRB   &   Exp. time (hr)   &            mag       &    mag error \\
  \hline
  1.23    &    0.08 &    17.32 &     0.05 \\
  1.34    &    0.08 &    17.42 &     0.05 \\
  1.44    &    0.08 &    17.50 &     0.05 \\
  1.54    &    0.08 &    17.56 &     0.05 \\
  1.65    &    0.08 &    17.70 &     0.05 \\
  1.76    &    0.08 &    17.76 &     0.05 \\
  1.86    &    0.08 &    17.81 &     0.06 \\
  1.96    &    0.08 &    17.96 &     0.05 \\
  2.07    &    0.08 &    17.92 &     0.05 \\
  2.17    &    0.08 &    18.03 &     0.05 \\
  2.28    &    0.08 &    18.12 &     0.05 \\
  2.38    &    0.08 &    18.19 &     0.07 \\
  2.48    &    0.08 &    18.31 &     0.08 \\
  2.59    &    0.08 &    18.17 &     0.13 \\
  \hline

\end{tabular}
\end{table}

\section{Equivalent widths}

In this appendix we present the rest-frame equivalent widths of the
most important transitions reported in Table 2. These equivalent
widths represent the sum of the contributions of all the three
components which model each transition (see Sect. 4.2).

\begin{table}[ht]
  \caption{Rest-frame equivalent widths (expressed in \AA) for 
some representative absorption features.}
\centering
\smallskip
\begin{tabular}{|l|c|c||l|c|c|}
  \hline
  Line                   &   EW    & $\Delta$EW &  Line       &    EW  & $\Delta$EW  \\
  \hline
\ion{C}{ii}~$\lambda$1334      &    0.99  &    0.04 &  \ion{C}{ii}~$\lambda$1335*  & 0.51 & 0.04 \\
\ion{C}{iv}~$\lambda$1548      &    0.75  &    0.04 &  \ion{C}{iv}~$\lambda$1550   & 0.66 & 0.04 \\
\ion{N}{i}~$\lambda$1134c      &    0.44  &    0.06 &  \ion{N}{v}~$\lambda$1238    & 0.11 & 0.04 \\
\ion{N}{V}~$\lambda$1242       &    0.06  &    0.04 &  \ion{O}{i}~$\lambda$1302    & 0.92 & 0.04 \\
\ion{O}{i}~$\lambda$1306*2     &    0.21  &    0.04 &  \ion{O}{vi}~$\lambda$1031   & 1.37 & 0.08 \\
\ion{Mg}{i}~$\lambda$2852      &    0.80  &    0.04 &  \ion{Mg}{ii}~$\lambda$2796  & 2.07 & 0.04 \\
\ion{Mg}{ii}~$\lambda$2803     &    1.97  &    0.04 &  \ion{Al}{ii}~$\lambda$1670  & 1.00 & 0.03 \\
\ion{Al}{iii}~$\lambda$1854    &    0.34  &    0.04 &  \ion{Al}{iii}~$\lambda$1862 & 0.18 & 0.04 \\
\ion{Si}{ii}~$\lambda$1526     &    1.07  &    0.05 &  \ion{Si}{ii}~$\lambda$1808  & 0.63 & 0.10 \\
\ion{Si}{ii}~$\lambda$1309*    &    0.30  &    0.04 &  \ion{Si}{iv}~$\lambda$1393  & 0.48 & 0.03 \\
\ion{Si}{iv}~$\lambda$1402     &    0.34  &    0.03 &  \ion{P}{i}~$\lambda$1674    & 0.03 & 0.02 \\
\ion{P}{ii}~$\lambda$1152      &    0.17  &    0.05 &  \ion{P}{v}~$\lambda$1117    & 0.15 & 0.08 \\
\ion{S}{ii}~$\lambda$1250      &    0.30  &    0.04 &  \ion{S}{ii}~$\lambda$1253   & 0.44 & 0.04 \\
\ion{S}{vi}~$\lambda$944       &    0.20  &    0.12 &  \ion{C}{ii}~$\lambda$1334   & 0.51 & 0.04 \\
\ion{Ar}{i}~$\lambda$1066      &    0.22  &    0.07 &  \ion{Ca}{i}~$\lambda$4227   & 0.16 & 0.06 \\
\ion{Ca}{ii}~$\lambda$3934     &    0.98  &    0.06 &  \ion{Ca}{ii}~$\lambda$3969  & 0.94 & 0.06 \\
\ion{Cr}{ii}~$\lambda$2056     &    0.35  &    0.03 &  \ion{Cr}{ii}~$\lambda$2066  & 0.27 & 0.03 \\
\ion{Fe}{ii}~$\lambda$1608     &    0.91  &    0.03 &  \ion{Fe}{ii}~$\lambda$2344  & 1.51 & 0.22 \\
\ion{Fe}{ii}~$\lambda$1618*    &    0.07  &    0.03 &  \ion{Fe}{ii}~$\lambda$2396* & 0.85 & 0.22 \\
\ion{Fe}{ii}~$\lambda$1629*2   &    0.06  &    0.03 &  \ion{Fe}{ii}~$\lambda$1636*3& 0.51 & 0.04 \\
\ion{Fe}{ii}~$\lambda$1642*3   &    0.04  &    0.03 &  \ion{C}{ii}~$\lambda$1334   & 0.51 & 0.04 \\
\ion{Fe}{ii}~$\lambda$2414     &    0.16  &    0.12 &  \ion{Fe}{ii}~$\lambda$1702*5& 0.19 & 0.03 \\
\ion{Fe}{ii}~$\lambda$1712*6   &    0.04  &    0.03 &  \ion{Fe}{ii}~$\lambda$1635*9& 0.04 & 0.03 \\
\ion{Ni}{ii}~$\lambda$1701     &    0.21  &    0.03 &  \ion{Ni}{ii}~$\lambda$1741  & 0.25 & 0.03 \\
\ion{Ni}{ii}~$\lambda$2217*2   &    0.41  &    0.05 &  \ion{Ni}{ii}~$\lambda$2223*2& 0.15 & 0.05 \\
\ion{Zn}{ii}~$\lambda$2026     &    0.34  &    0.03 &                              &      &      \\
 
  \hline

\end{tabular}
\end{table}

\end{document}